\pdfoutput=1   
\documentclass[10pt]{article}
\usepackage{cancel}
\usepackage{graphicx}
\usepackage[colorlinks=true, pdfstartview=FitV, linkcolor=blue, citecolor=blue,urlcolor=blue]{hyperref}
 
\usepackage{framed,color}
\definecolor{shadecolor}{rgb}{0.90, 0.90, 0.90}

\usepackage{amsbsy}
\usepackage{amssymb,amsmath}
 \usepackage{amsfonts}
\usepackage{graphicx}

\usepackage{cancel}
\usepackage{mathtools}
\usepackage{amsmath}
\usepackage{amsthm}
\usepackage{amsfonts}
\usepackage{amssymb}
\usepackage{mathrsfs}
\usepackage{graphicx}
\usepackage[T1]{fontenc}
\usepackage[OT2,T1]{fontenc}
\usepackage[latin1]{inputenc}
 \usepackage{lmodern}
\usepackage[all,cmtip]{xy}

\usepackage{bbm}
\usepackage[vcentermath]{youngtab}
\usepackage{calc}

\usepackage{setspace}
\usepackage{epigraph}
\usepackage{mathdots}
\usepackage{mathabx}
\usepackage{braket}
\usepackage{longtable}
\usepackage{float}

\usepackage{imakeidx}
\makeindex[columns=2]
\def\eqref#1{(\ref{#1})}
\newtheorem{theorem}{Theorem}[section]
\newtheorem{example}{Example}[section]
\newtheorem{exercise}{Exercise}[section]

\newtheorem{lemma}{Lemma}[section]
\newtheorem{remark}{Remark}[section]

\newtheorem{proposition}{Proposition}[section]
\newtheorem{corollary}{Corollary}[section]
\newtheorem{definition}{Definition}[section]

\def\bre{\begin{remark}}
\def\ere{\end{remark}}
\def\bth{\begin{theorem}}
\def\eth{\end{theorem}}
\def\bcr{\begin{corollary}}
\def\ecr{\end{corollary}}
\def\bex{\begin{example}\small}
\def\eex{\end{example}}
\def\bexr{\begin{exercise}\small}
\def\eexr{\end{exercise}}
\def\ble{\begin{lemma}}
\def\ele{\end{lemma}}

\def\bde{\begin{definition}}
\def\ede{\end{definition}}
\def\bpr{\begin{proposition}}
\def\epr{\end{proposition}}
\def\be{\begin{equation}}
\def\ee{\end{equation}}
\def\bea{\begin{eqnarray}}
\def\eea{\end{eqnarray}}
\def\beas{\begin{eqnarray*}}
\def\eeas{\end{eqnarray*}}
\newlanguage\fakelanguage
\newcommand\cyr{\fontencoding{OT2}\fontfamily{wncyr}\selectfont
   \language\fakelanguage}
\DeclareTextFontCommand{\textcyr}{\cyr}

\numberwithin{equation}{section}

\numberwithin{theorem}{section}

\numberwithin{proposition}{section}

\numberwithin{definition}{section}

\numberwithin{remark}{section}

\numberwithin{lemma}{section}

\numberwithin{corollary}{section}

\date{}
\begin{document}

\centerline{\bf \Large Generalized Hermite Polynomials and  }
\vskip 0.3 cm 
\centerline{\bf \Large   Spectral Degeneracies of a Singular Sextic Oscillator}

\vskip 0.5 cm 
\centerline{\Large Davide Guzzetti \& Dmitrii Rachenkov}

\vskip 0.2 cm 
\centerline{SISSA, 
Via Bonomea 265, 34136 Trieste, Italy}
\vskip 0.2 cm 
\centerline{ INFN Sezione di Trieste, 
via Valerio 2, 34127 Trieste, Italy}

\begin{abstract}

We study a quasi-exactly solvable singular sextic oscillator and its algebraic spectrum. For a distinguished range of  parameters, we prove that the discriminant of the characteristic polynomial of the matrix determining the algebraic spectrum admits a natural  factorization into three factors. One of these factors is the square of a generalized Hermite polynomial $H_{mn}$, whose zeros are poles of a rational solution of the fourth Painlev\'e equation.
 Hence, the spectral degeneracies (level crossing points) corresponding to a component of the discriminant locus are in exact correspondence with the zeros of generalized Hermite polynomials, providing an {\it exact} Painlev\'e IV analogue of the Shapiro--Tater {\it asymptotic} correspondence originally conjectured  for the quartic oscillator and Painlev\'e II. We also characterize the values of the parameters for which the sextic oscillator admits simultaneously two quasi-polynomial eigenfunctions with opposite exponential behaviour at infinity, and show that this phenomenon is also governed by generalized Hermite polynomials.

Our result also yields a new determinantal representation of $H_{mn}$ as the resultant of the characteristic polynomials of two complementary blocks of the  matrix determining the algebraic spectrum.
\end{abstract}

\tableofcontents

\vskip 1 cm 
\noindent{\bf Notation}  The letter $i$ will be used only to denote $\sqrt{-1}$. 
The set of natural numbers $\mathbb{N}$ contains $0$, that is  $\mathbb{N}=\{0,1,2,3,\dots\}$. If $\mathcal{M}$ is a matrix of dimension $n$, $I_n$ the identity matrix of dimension $n$  and $\lambda\in\mathbb{C}$, we will often write 
$\mathcal{M}+\lambda$ for the matrix $\mathcal{M}+\lambda I_n$.

\section{Introduction}

\subsection{The problem}
We study the spectral problem
\be
\label{30aprile2024-2}
-\frac{d^2y}{dx^2}+\left(x^6+2bx^4+(b^2-2M-3)x^2+\frac{\gamma}{x^2}\right)y=\lambda y, \quad\quad M,b,\gamma,\lambda\in\mathbb{C},
\ee
for  a sextic anharmonic oscillator with an inverse square singularity (the so called centrifugal term). Such operator belongs to the class of quasi-exactly solvable Schr\"odinger operators \cite{ST,Ush,Turb-1}: for special values of the parameters, part of the spectrum can be described algebraically as the spectrum of a finite dimensional matrix, and the eigenfunctions are exactly computed.

For the sextic oscillator without the singular term, corresponding to $\gamma=0$, the algebraic spectrum is known to exhibit degeneracies for special values of the parameter $b$, sometimes  called level-crossing points. A striking property, well observed for the quartic anharmonic oscillator \cite{ST-1,BCG-1}, is that the distribution of these exceptional values may  be asymptotically related  to that of  the zeros and poles  of rational solutions of certain  Painlev\'e equations.

The aim of this paper is to investigate the  phenomenon for the singular sextic oscillator, where $\gamma$ is allowed to be non-zero. In this setting the algebraic spectrum is associated to quasi-polynomial eigenfunctions with a prescribed exponential behaviour at infinity, and is given by a finite matrix depending on $b$. The spectral degeneracies corresponds to the values of $b$ where the discriminant of the characteristic polynomial vanishes. 

We show that, for suitable half-integer values of $M$ and suitable $\gamma$, the discriminant admits a natural factorization into three factors, one of which is the square of a generalized Hermite polynomial. The relation between the exceptional values of $b$ and  the zeros $a$  of generalized Hermite polynomials is not merely {\it asymptotic}; they {\it exactly  coincide} after the simple rescaling $a=b/\sqrt2$.  

It is well known that the Painlev\'e IV equation admits a family of rational solutions that can be expressed as ratios of generalized Hermite polynomials. In view of this fact, our factorization of the discriminant highlights an exact relationship between a component of the zero-locus of the discriminant of \eqref{30aprile2024-2} and the distribution of the poles and zeros of the rational solutions of the Painlev\'e  IV equation.

The above result  can be regarded as an analogue of a conjecture formulated by B. Shapiro and M. Tater in \cite{ST-1},  regarding the quartic oscillator and Painlev\'e II. 
  The quartic oscillator   depends on a parameter, analogous to our $b$. The  spectral problem is reduced to an eigenvalue problem for a finite dimensional  matrix, giving  the algebraic spectrum, and the  discriminant of the characteristic polynomial is a polynomial in the parameter.   Shapiro an Tater conjectured that  the roots of the discriminant and the roots of the  Vorob'ev-Yablonsky polynomials   {\it asymptotically} form  two coinciding lattices as the number of roots tends to infinity in an appropriate way.   This is equivalent to conjecturing an asymptotic correspondence between the roots of the discriminant and  the distribution of zeros and poles of the  rational solution of  Painlev\'e II, well known to be  a ratio of  Vorob'ev-Yablonsky polynomials. This conjecture for Painlev\'e II was essentially  proved in \cite{BCG-1}. 

In this paper, we also characterize the values of the parameters for which the sextic oscillator admits simultaneously two linearly independent quasi-polynomial eigenfunctions with opposite exponential behaviour at infinity, and show that this phenomenon is also governed by generalized Hermite polynomials.

\vskip 0.2 cm 
Anharmonic oscillators have been extensively studied in the literature. They have gained particular interest in connection with the ODE/IM correspondence \cite{D1..?,BazZ,Suzz,DDT,MRV,MRV2,Deg1,Deg2},  Painlev\'e equations and isomonodromy deformations \cite{DMTritr-1,DMTritr-2, MR, MR1,Maso-Bridgl-1}. In particular, the  sextic oscillator was introduced  in   \cite{Singh} with $\gamma=0$. Special quasi-polynomial eigenfunctions given by a polynomial times the exponential factor $\exp\{-x^4/4-bx^2/2\}$  and their algebraic spectrum where studied in \cite{Turb-2}. Properties of eigenfunctions were also studied in \cite{Erem-Shap}. Further spectral properties where investigated in \cite{Turb-1,Ush,BenD-1,BenD-2}. In \cite{ST-2}, asymptotic properties of the algebraic spectrum (always for $\gamma=0$) where studied, and  the problem was posed for the distribution of the values $b$, the above mentioned  level crossing points, such that the cardinality of the algebraic spectrum is less than expected.

 \subsection{Outline of the Results} 
\label{28giugno2026-1}

\begin{figure}
\centerline{\includegraphics[width=0.6\textwidth]{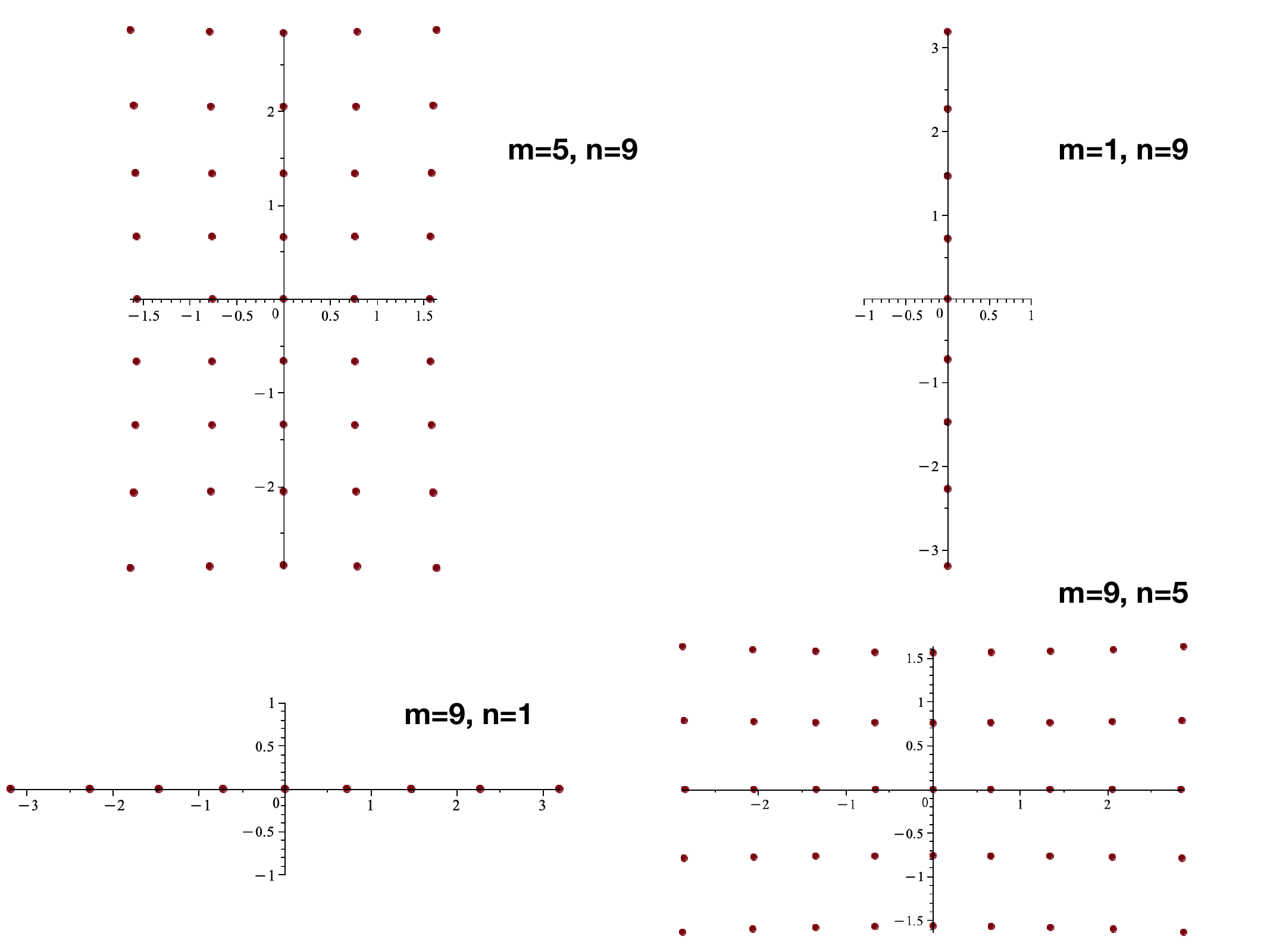}}
\caption{Roots of the generalized Hermite polynomial for several values of $m,n$.}
\label{14-giugno-2025-2}
\end{figure}

The Painlev\'e equation IV
$$
\frac{d^2 u}{dt^2}= \frac{1}{2u}\left(\frac{du}{dt}\right)^2+\frac{3}{2}\,u+4 t\,u^2+2(t^2+1-2\theta_\infty)u-\frac{8\theta_0^2}{u}
$$
 admits rational solutions if and only if $\theta_0$ and $\theta_\infty$ take some special rational values depending on  a pair of integers $(m,n)\in\mathbb{Z}^2$. These solutions are given by the ratio of either  generalized Hermite polynomials or  the generalized Okamoto polynomials \cite{NY}, so that their zeros and poles depend on the zeros of the polynomials. 
In this paper, we consider the generalized Hermite case. A  generalized Hermite polynomial  $H_{mn}(t)$, defined for integers $m\geq 0$, $n\geq 0$  (see Section \ref{15giugno2026-3}), has degree $m\cdot n$ and simple roots, typically distributed as in Figure \ref{14-giugno-2025-2}. They form an almost rectangular lattice,  with $m$ points on the basis and $n$ on the height, precisely characterized in  \cite{MR, MR1} (see also \cite{BM}). 

We are going to show that the above roots, responsible for the zeros and poles of the associated rational solutions of Painlev\'e IV, also have a deep connection with the spectral problem \eqref{30aprile2024-2}. 

\vskip 0.2 cm 
  Let 
  \be 
\label{17febbraio2026-2}
\vartheta(x):=\frac{x^4}{4}+\frac{b\,x^2}{2}.
\ee It is relatively simple to prove that  problem  \eqref{30aprile2024-2} admits a solution $(\Lambda, \,y_1(x,\Lambda))$ with eigenvalue $\lambda=\Lambda$ and   a {\it quasi-polynomial} eigenfunction\footnote{It can be written as a polynomial of degree $2N$ times $x^{M-2N} \exp\{-\vartheta(x)\}$.
 }  
 $$
 y_1(x,\Lambda)= \left(\sum_{k=0}^N c_{2k}(\Lambda)\,x^{-2k} \right)\, x^M\, \exp\{-\vartheta(x)\},\quad c_0\neq 0,\quad N\in\mathbb{N},
$$
if and only if (see Proposition \ref{13aprile2023-4})
\be
\label{15giugno2026-5}
\gamma=(2N-M+1)(2N-M),
\ee
 for some $N\in\mathbb{N}$, and $-\Lambda$ is eigenvalue of a certain  $(N+1)\times(N+1)$  matrix $\mathcal{M}=\mathcal{M}(b,M,N)$, that will be explicitly given in Section \ref{17febbraio2026-1}. 
 The matrix $\mathcal{M}$ naturally appears  when solving the recurrence relations for the coefficients $c_{2k}$, which form  an eigenvector of  the matrix relative to the eigenvalue $-\Lambda$. 
 \vskip 0.2 cm 
 For fixed $b$, $M$ and $N$, the eigenvalues of $\mathcal{M}$ are called  {\it algebraic} spectrum and the eigenvalue problem \eqref{30aprile2024-2} is said to be {\it exactly solvable}. The cardinality of the algebraic spectrum for fixed $M$, $N$ and generic $b$ is $N+1$. 
 \vskip 0.2 cm 
 
\begin{figure}
\centerline{\includegraphics[width=1.2\textwidth]{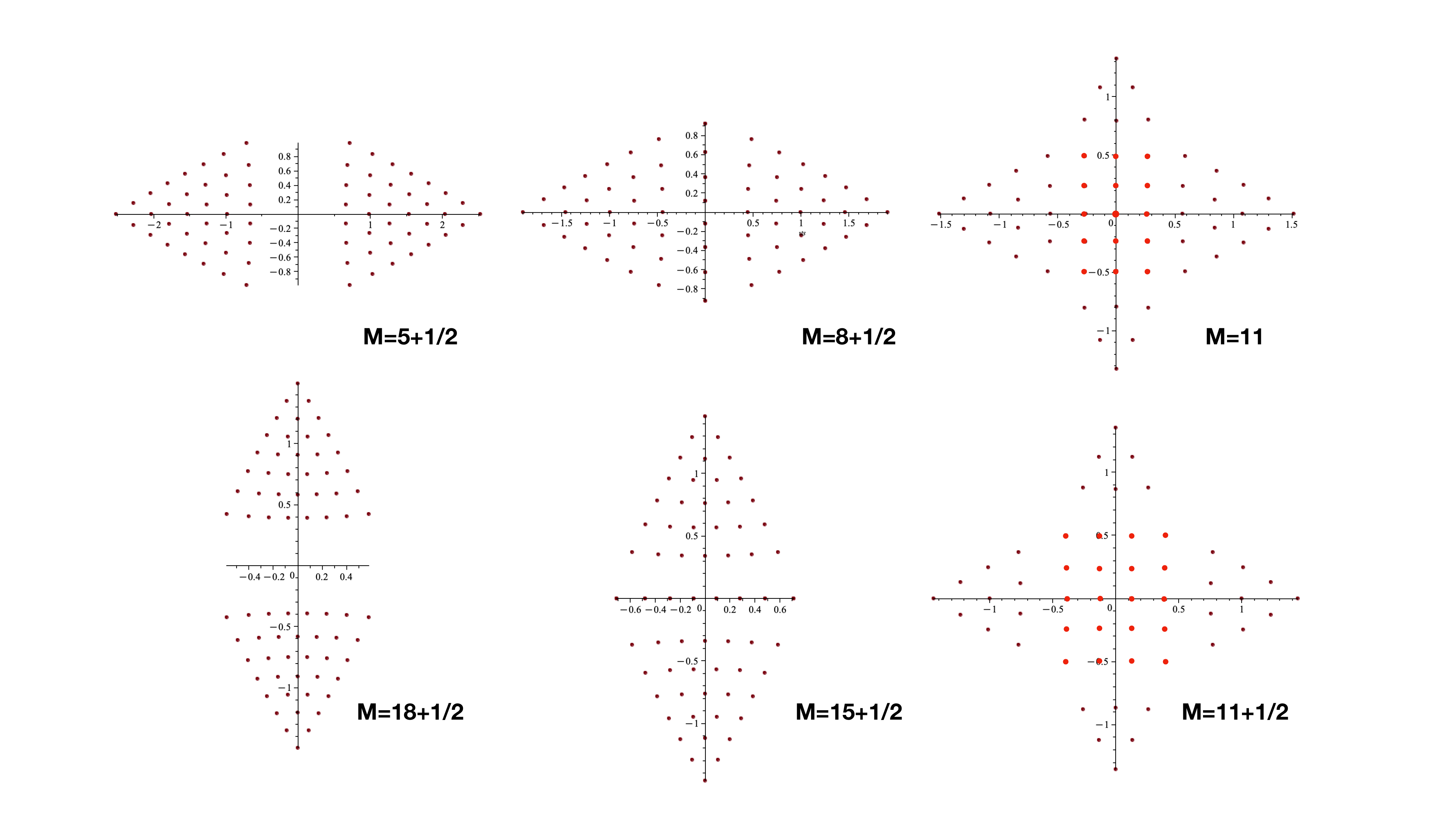}}
\caption{Roots of resultant for $N=8$ and several values of $M$ half integer or integer. }
\label{21marzo2026-1}
\end{figure}

  The cardinality of the spectrum becomes less than $N+1$ if 
an element $\Lambda$ has  algebraic multiplicity greater than one, as eigenvalue of $-\mathcal{M}$. This happens when  $b$ is a  root of the discriminant of the characteristic polynomial 
 \be
\label{21febbraio2026-1}
p(\lambda;\,b):=\det(\mathcal{M}(b,M,N)+\lambda I_{N+1}),
 \ee
  or equivalently a root of the polynomial  in  $b$ given by 
\be
\label{17febbraio2026-3}
{\rm res}_\lambda\bigl(p(\lambda;\,b),\partial_\lambda p(\lambda;\,b)\bigr),
\ee
 where the symbol $ {\rm res}_\lambda$ stands for the resultant or $p$ and $\partial_\lambda p$ with respect to the variable $\lambda$.  Representing these roots as points in the complex plane, when $M\in\mathbb{R}$ we obtain a distribution invariant by reflection w.r.t. the horizontal and vertical axes (as proved in  Section \ref{21febbraio2026-3}), with a characteristic shape, as represented  for $M$ integer or half integer in Figure \ref{21marzo2026-1}.

\begin{figure}
\centerline{\includegraphics[width=0.65 \textwidth]{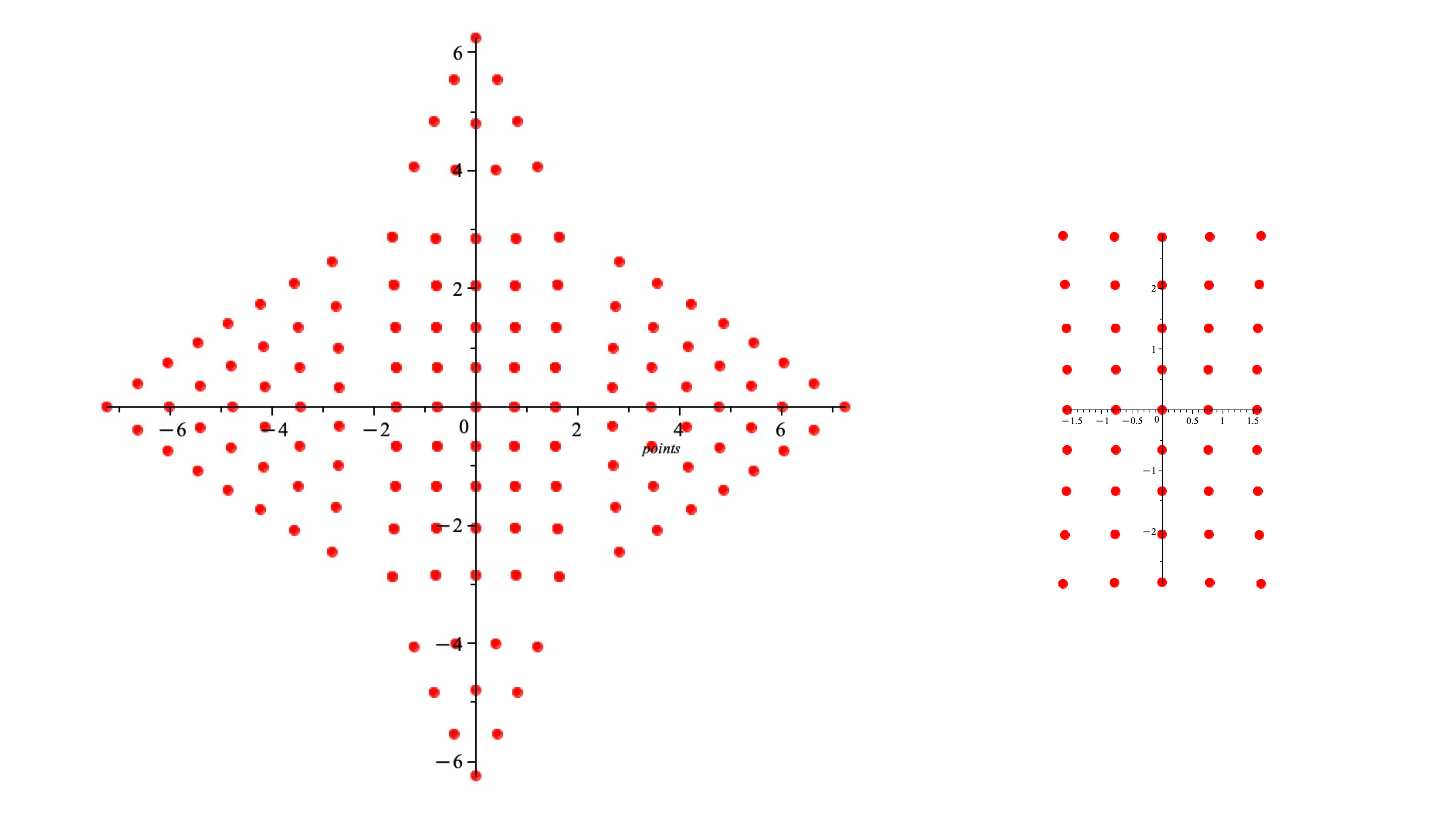}}
\caption{Left side: roots of the resultant of the sextic oscillator in the $a$-plane, where $a:=b/\sqrt{2}$, for  $(M,N)=(17+1/2,~13)$, or equivalently $(m,n)=(5,9)$, where we define $m := M - N + 1/2$, $
n := 2N - M + 1/2$. Numerical evidence is that the central rectangle has base with $m$ dots and height with $n$, and exactly overlaps with the set of roots of the Hermite polynomial $H_{m,n}(a)$, represented in the the right side of the figure for $m=5,n=9$.}
\label{14-giugno-2025}
\end{figure}

\vskip 0.2 cm 
We   restrict now to the case of half-integer $M$ with the additional constraint 
\be
\label{16marzo2026-1}
 M\in \mathbb{N}_{\geq 1}+\frac{1}{2},
 \quad
  \hbox{ and }
 \quad
 \frac{M}{2}+\frac{1}{4}\leq N\leq M-\frac{1}{2}.
\ee
We define the positive integers
\be
\label{2giugno2026-5}
m: = M - N + \frac{1}{2},\quad\quad 
n := 2N - M + \frac{1}{2}.
\ee
In this case, the distribution of the roots $b$ of \eqref{17febbraio2026-3} in the complex plane exhibits a central part  appearing as  an  approximately rectangular lattice, consisting of $m$ points along its base and $n$ points along its height,   as  in Figure \ref{14-giugno-2025} (left part) and   in Figure \ref{21marzo2026-1} for the cases  $M=8+1/2$, $M=11+1/2$, $M=15+1/2$.\footnote{We also represent  in Figure  \ref{21marzo2026-1} the case $M=11$. For integer $M$, with $M$ and $N$ is a suitable range, the distribution of the roots of the discriminant also displays  a central part with a rectangular shape, but we will not consider the case of $M$ integer in this paper, which is  different in an essential way and may be asymptotically related to the zeros of generalized Okamoto polynomials. No natural factorization \eqref{15giugno2025-2S-BIS} is possible for integer $M$.}

The crucial observation for our result is now the following. 
Let  $a\in\mathbb{C}$ denote a root of $H_{mn}(t)$, that is $H_{mn}(a)=0$. 
Numerical computations indicate that these roots coincide with the roots of the resultant \eqref{17febbraio2026-3} belonging to the central almost rectangular lattice mentioned above, upon identifying $a=b/\sqrt{2}$ (see   Figure \ref{14-giugno-2025}).  This means that each root of $H_{mn}(t)$, which is simple, overlaps with exactly one point representing a (possibly repeated) root of \eqref{17febbraio2026-3} in the central almost rectangular lattice, and all points of the central lattice are accounted for in this way.\footnote{The algebraic multiplicity of the corresponding roots of the resultant is not known a priori; we shall show later that it is equal to two.}

The above observation is an  analogue of the  Shapiro-Tater's conjecture mentioned in the introduction. We characterize and prove this  observation  as an {\it exact result}, namely  for {\it all values} of $m,n \geq 1$, and not only in an {\it asymptotic limit} as was the case of \cite{ST-1,BCG-1}. This is the content of the main theorem of the paper:

\vskip 0.2 cm 
\bth
\label{17giugno2025-1}
 Consider the sextic oscillator \eqref{30aprile2024-2}, with $\gamma$ as in \eqref{15giugno2026-5},  
and satisying the constraint \eqref{16marzo2026-1}.  Then, the resultant in \eqref{17febbraio2026-3},  as a polynomial in the variable $a=b/\sqrt{2}$, admits  the  factorization
 \be
 \label{15giugno2025-2S-BIS}
 {\rm res}_\lambda\bigl(\,p(\lambda;\,\sqrt{2}a),\,\partial_\lambda p(\lambda;\,\sqrt{2}a)\,\bigr)=
 (-1)^{mn}c_{mn}^{-1}\,\,  r_1(a) \, r_2(a) \, H_{mn}(a)^2,
 \ee
 where
 \begin{itemize}
 \item  $c_{mn}\in\mathbb{C}\backslash\{0\}$ is a constant,  
 \item $H_{mn}(a)$ is the generalized Hermite polynomial relative to  $m$ and $n$ defined in \eqref{2giugno2026-5}, 
 \item  
 $r_1(a)$ and $r_2(a)$ are the following  polynomials, of degrees $n(n-1)$ and $m(m-1)$  respectively,
  $$ 
 r_1(a):={\rm res}_\lambda\left( \det
 \left(\mathcal{M}_1(\sqrt{2}a,M,N)+\lambda I_n\right),~ \partial_\lambda\det\left(\mathcal{M}_1(\sqrt{2}a,M,N)+\lambda I_n\right)\right),
 $$
 $$ 
 r_2(a):={\rm res}_\lambda\left( \det\left(\mathcal{M}_2(\sqrt{2}a,M,N)+\lambda I_m\right),~ \partial_\lambda\det\left(\mathcal{M}_2(\sqrt{2}a,M,N)+\lambda I_m\right)\right),
 $$
 where $\mathcal{M}_1$ and  $\mathcal{M}_2$ are, respectively, the upper-left $n\times n$ block and the lower-right $m \times m$ block of $\mathcal{M}(b,M,N)$. 
 \item 
 The generalized Hermite polynomial is, up to the constant $c_{mn}$, represented as a resultant
$$ 
 \begin{aligned}
 H_{m,n}(a)= c_{mn}{\rm res}_\lambda\left( \det\left(\mathcal{M}_1\bigl(\sqrt{2}a,M,N\bigr)+\lambda I_n\right), \,\det\left(\mathcal{M}_2\bigl(\sqrt{2}a,M,N\bigr)+\lambda I_m \right)\right),
 \\
M=2m + n - \frac{3}{2},\quad N=m+n-1.
\end{aligned} 
$$
 \end{itemize}
\eth

\begin{figure}
\centerline{\includegraphics[width=0.40\textwidth]{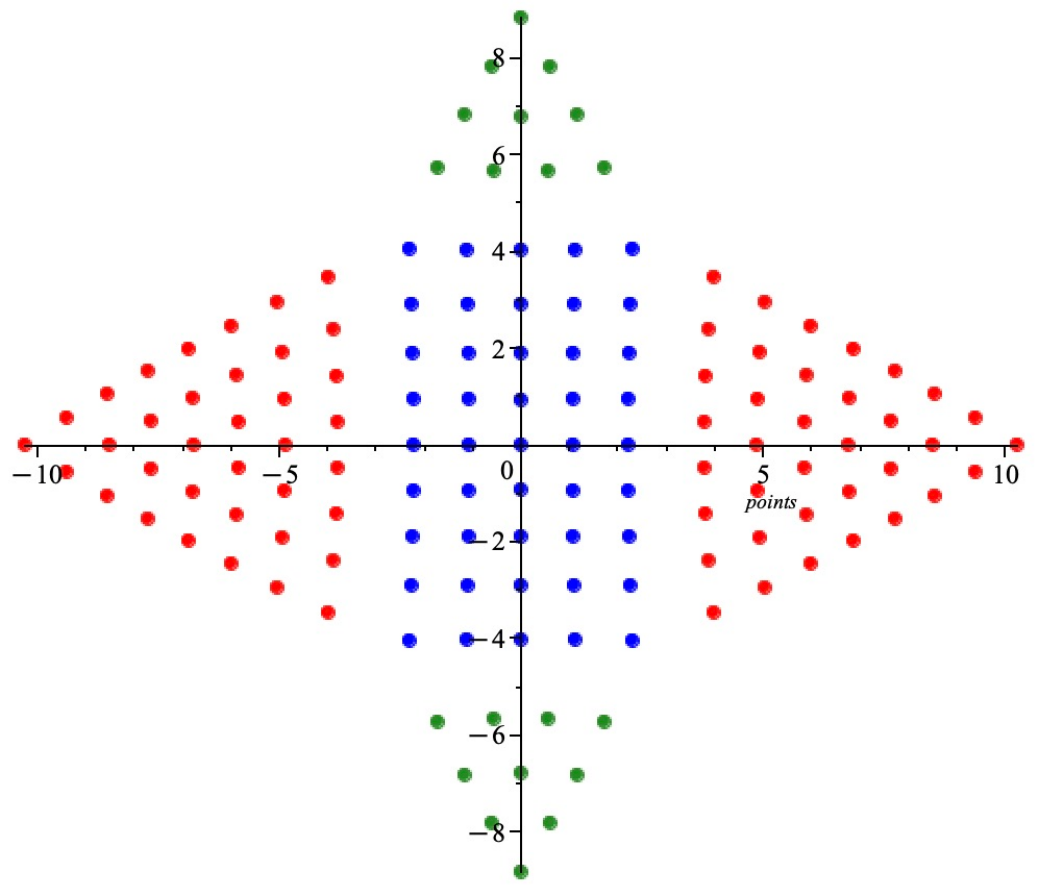}}
\caption{The roots of the polynomials in the factorization \eqref{15giugno2025-2S-BIS}, for   $(m,n)=(5,9)$, are represented in the $b$-plane in different colours corresponding to different factors: red for $r_1$, green for $r_2$ and blue for $H_{mn}$. Each blue dot is  a double root of ${\rm res}_\lambda\left(\det(\mathcal{M}(\sqrt{2}a)+\lambda),~ \partial_\lambda\det(\mathcal{M}(\sqrt{2}a)+\lambda)\right)$.}
\label{15giugno2025-1}
\end{figure}

\vskip 0.2 cm 
The roots of $r_1$, $r_2$ and $H_{m,n}$  are all symmetric by reflection through the horizontal and vertical axes (Section \ref{21febbraio2026-3}).  
Numerical computations clarify that the  factorization \eqref{15giugno2025-2S-BIS} above is responsible for the ``star shaped'' distribution of the zero of the resultant, the central rectangular part being due to the Hermite polynomial. See Figure \ref{15giugno2025-1}.

\vskip 0.2 cm 
We also prove a second  facet of the correspondence between the sextic oscillator and Hermite-type rational solutions of Painlev\'e IV.
Preliminarily, we observe that the sextic oscillator  also admits a {quasi-polynomial solution} with positive exponential factor
$$
y_2(x,\Lambda):=\left(\sum_{k=0}^N d_{2k}(\Lambda) x^{-2k}\right)\,x^{-M-3}\,  \exp\{\vartheta(x)\},\quad d_0\neq 0,\quad  N\in\mathbb{N},
$$
if and only if $\gamma=(2N+M+4)(2N+M+3)$ and $\Lambda$ is a root of 
\be
\label{21febbraio2026-2}
\widetilde{p}(\lambda,b):=\det\Bigl(i\mathcal{M}\bigl( i b,\, -M-3,\,N\bigr)+ \lambda\Bigr)=0 ,
\ee
Also in this case, the (possibly repeated)  roots of $\widetilde{p}(\lambda,b)$ above  are called {\it  algebraic spectrum} and the eigenvalue problem \eqref{30aprile2024-2} is said to be {\it exactly solvable}. 

\vskip 0.2 cm 

One can impose the condition that   $\gamma$ is simultaneously equal to $(2N_1-M+1)(2N_1-M)$ and  $(2N_2+M+4)(2N_2+M+3)$ for some integers $N_1$ and $N_2$, in order to investigate the simultaneous existence of quasi-polynomial solutions with negative and positive exponentials.  The condition of existence of two quasi-polynomial solutions is proved in the following proposition, which represents a  second  facet of the correspondence between the sextic oscillator and Hermite-type rational solutions of Painlev\'e IV. 
\vskip 0.3 cm 
\noindent
{\bf Proposition  [Proposition \ref{8febbraio2026-5}]}
{\it  The eigenvalue problem  \eqref{30aprile2024-2} for the sextic oscillator has   simultaneously  two quasi-polynomial  eigenfunctions $y_1(x,\Lambda) $ and $y_2(x,\Lambda)$  with negative and positive exponential and    $N_1+1$ and $N_2+1$ terms  respectively, if and only if the following three conditions hold:

\begin{itemize}
\item[1)] The coefficients of \eqref{30aprile2024-2} take the values 
\be
\label{15marzo2026-1}
M=N_1-N_2-\frac{3}{2}, \quad 
\quad 
 \gamma= \Bigl(N_1 + N_2 + \frac{3}{2}\Bigr) \Bigl(N_1 + N_2 + \frac{5}{2}\Bigr) ;
 \ee
  \item[2)] 
   $a:=b/\sqrt{2}$ is a  root of the generalized Hermite polynomial 
 $$H_{N_2+1, N_1+1}(a)=0;
 $$
\item[3)] $\lambda=\Lambda$, where   $-\Lambda$ is common eigenvalue of the matrices 
 $$ 
\mathcal{M}\Bigl(b,N_1-N_2-\frac{3}{2},N_1\Bigr) \quad\hbox{ and } \quad \quad   i\mathcal{M}\Bigl(i b,N_2-N_1-\frac{3}{2},N_2\Bigr).
 $$
 \end{itemize}
}

\begin{figure}
\centerline{\includegraphics[width=0.37\textwidth]{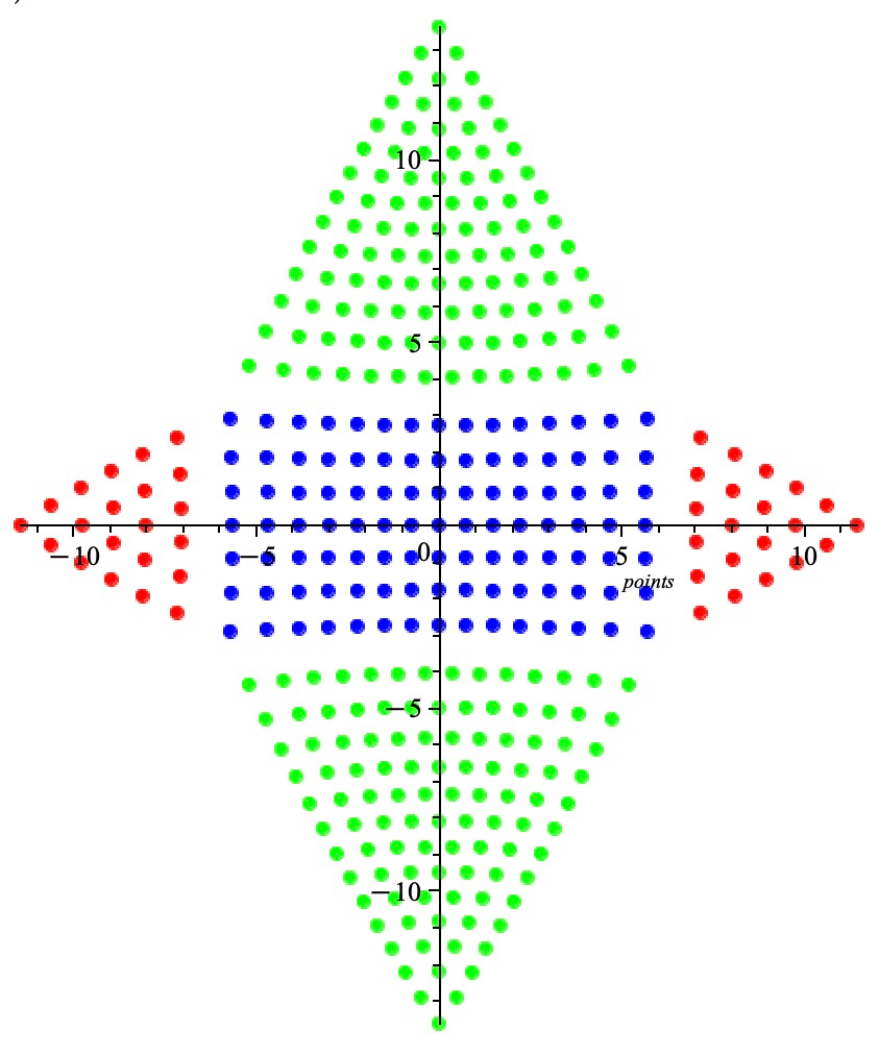}}
\caption{Case $m:=N_2+1=15$, $n:=N_1+1=7$. In red, the roots of $R_1$, in green the roots of $R_2$, in blue the roots of $R_3$.  A red or green dot means that for that value of $b$ the sextic oscillator has a repeated eigenvalue with  a quasi-polynomial solution with negative exponent or  with positive exponent respectively. A blue dot means that for that value of $b$ the sextic oscillator has a simple  eigenvalue with simultaneously  two quasi-polynomial solutions with negative and positive exponent.}
\label{21febbraio2026-4}
\end{figure}

\begin{figure}
\centerline{\includegraphics[width=0.65\textwidth]{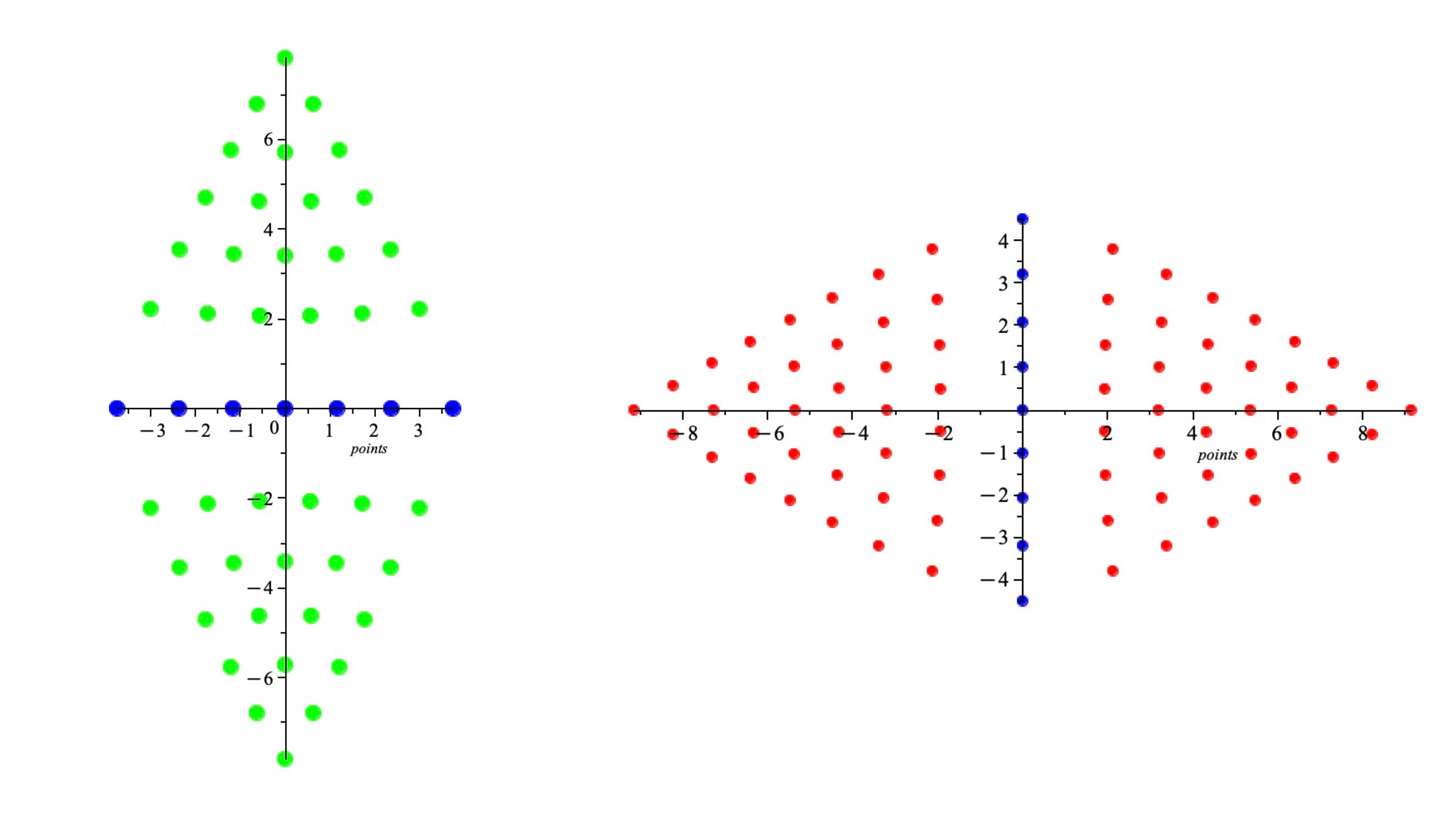}}
\caption{Left figure: the case $m:=N_2+1=7$, $n:=N_1+1=1$. In  green the roots of $R_2$, in blue the roots of $R_3$,  while $R_1$ is constant because  $N_1=0$. Right figure: the case $m:=N_2+1=1$, $n:=N_1+1=9$. In red, the roots of $R_1$,  in blue the roots of $R_3$, while $R_2$ is constant because $N_2=0$.}
\label{new-figure}
\end{figure}

Note that for $M$ and $\gamma$ in 
\eqref{15marzo2026-1},   the assumptions of Theorem \ref{17giugno2025-1}  do not hold. However,   also in this case we can distinguish three polynomials $R_1$, $R_2$ and $R_3$ which are analogous  to  $r_1$, $r_2$ and $H_{mn}$ of  Theorem \ref{17giugno2025-1}.  
  With  paramaters \eqref{15marzo2026-1}, let $p(\lambda; b, M,N_1)$ be the polynomial defined in \eqref{21febbraio2026-1} for $N=N_1$, and   $\widetilde{p}(\lambda; b, M,N_2)$ be the polynomial define in \eqref{21febbraio2026-2} for $N=N_2$.   
  Then, we define the polynomials $R_k$ as follows.  \begin{itemize}
   \item
 $
  R_1(b, N_1,N_2):= {\rm res}_\lambda( p(\lambda; b, M,N_1),\partial_\lambda p(\lambda; b, M,N_1));
   $
   
  if $b$ is a root, the  quasipolynomial solution $y_1(x,\Lambda) $  is associated with a repeated eigenvalue $-\Lambda$ of $\mathcal{M}\Bigl(b,N_1-N_2-\frac{3}{2},N_1\Bigr)$. 
   The roots visually appear to form two  almost exagonal lattices as the red dots in Figures \ref{21febbraio2026-4} and \ref{new-figure}.

   \item 
   $
   R_2(b, N_1,N_2):= {\rm res}_\lambda( \widetilde{p}(\lambda; b, M,N_2),\partial_\lambda \widetilde{p}(\lambda; b, M,N_2))
   $;
   
    if $b$ is a root, the  quasipolynomial solution $y_2(x,\Lambda) $  with $N_2+1$ is associated with a repeated eigenvalue $-\Lambda$ of $ i\mathcal{M}\Bigl(i b,N_2-N_1-\frac{3}{2},N_2\Bigr)$. 
      The  roots visually appear to form two  almost exagonal lattices as  the green dots in Figure \ref{21febbraio2026-4} and   \ref{new-figure}.

   \item 
   $
   R_{3}(b, N_1,N_2):={\rm res}_\lambda(p(\lambda; b, M,N_1),\widetilde{p}(\lambda; b, M,N_2));
   $
   
    if $b$ is a root, the corresponding Schr\"odinger  equation with that $b$ and paramaters \eqref{15marzo2026-1} has  simultaneously two quasi-polynomial  eigenfunctions. 
    There is a non-zero constant $c_{N_2,N_1}$ such that  $R_3(b)=c_{N_2,N_1}\cdot H_{N_2+1, N_1+1}(b/\sqrt{2})
 $. The roots appear as the  blue dots  in Figures \ref{21febbraio2026-4} and \ref{new-figure} 

   \end{itemize}
  
Figure 
 \ref{21febbraio2026-4} (or the limiting case as in Figure \ref{new-figure}) has the  same structure of Figure  \ref{15giugno2025-1}, but the meaning is different. In Figure  \ref{15giugno2025-1}, all the roots are associated to spectral degeneracy for an eigenfunctin with given exponential behaviour at $\infty$, while in Figure  \ref{21febbraio2026-4} the central part is associated to the existence of two simultaneous eigenfunctions with opposite exponential behaviour at $\infty$, while the remaining roots are associated with spectral deceneracy with one eigenfunction  of either positive exponential behaviour (green dots) or  negative (red dots). The generalized Hermite polynomials are responsible for the roots in the central part of both Figures  \ref{15giugno2025-1} and  \ref{21febbraio2026-4}.
 
% \begin{figure}
% \centerline{\includegraphics[width=0.22\textwidth]{m=7-n=1.pdf}}
% \caption{Case $m:=N_2+1=7$, $n:=N_1+1=1$. In  green the roots of $R_2$, in blue the roots of $R_3$. Here $N_1=0$ and $R_1$ is constant.}
% \label{21febbraio2026-5}
% \end{figure}

% \begin{figure}
% \centerline{\includegraphics[width=0.4\textwidth]{m=1-n=9.pdf}}
% \caption{Case $m:=N_2+1=1$, $n:=N_1+1=9$. In red, the roots of $R_1$,  in blue the roots of $R_3$.  Here $N_2=0$ and $R_2$ is constant.}
% \label{21febbraio2026-6}
% \end{figure}

\subsection*{Acknowledgements} 
We would like to thank Marco Bertola and Tamara Grava for bringing our attention to this problem and for several discussions.

%%%%%%%%%%%%%%%%%%
\section{The Sextic Oscillator}
\label{17febbraio2026-1}

The aim of this section is to characterize quasi-polynomial solutions and  general solutions of \eqref{30aprile2024-2} with a canonical asymptotic behaviour.

\subsection{Quasi-polynomial solutions with negative exponential}
We recall the definition of the $\vartheta(x)$ in \eqref{17febbraio2026-2}, and  define  an $(N+1)\times(N+1)$ tridiagonal matrix $\mathcal{M}(b,M,N)$ with 
 entries
  \be
  \label{28maggio2024-1}
  \begin{aligned}
&\mathcal{M}_{jk}(b,M,N)
\\
\noalign{\medskip}
&:=\delta_{jk}\,b\,(4N-2M+3-4j)-2j\,(4N-2M+1-2j)\delta_{j,k-1}+4(N-j+2)\delta_{j,k+1},
\end{aligned}
\ee
for $j,k=1,\dots,N+1$, and $N\in\mathbb{N}$.

\bpr
\label{13aprile2023-4}
The eigenvalue problem   \eqref{30aprile2024-2}  has a solution $(\Lambda, y_1(x,\Lambda))$  of the  form 
$$
y_1(x,\Lambda):=Q_1(x,\Lambda) e^{-\vartheta(x)},
$$
where
$$
Q_1(x,\Lambda):=\Bigl(\sum_{k=0}^N c_{2k}(\Lambda)\, x^{-2k} \Bigr)\, x^M,\quad c_0\neq 0,\quad N\in\mathbb{N}.
$$ 
if and only if the following two conditions hold. \begin{itemize}

\item[$a)$] 
 For some $N\in\mathbb{N}$, the parameter $\gamma$ is 
$$\gamma=(2N-M+1)(2N-M);$$

\item[$b)$] $\Lambda$ is a root of \be
\label{25maggio2024-1}
\det(\mathcal{M}(b,M,N)+\lambda)=0 \quad \quad \hbox{ (degree $N+1$).}
\ee
\end{itemize}

\noindent
 The coefficients are  given by the formula 
  \be
\label{27aprile2026-2}
c_{2k}(\Lambda)=c_0\, \frac{(-1)^k\,\chi_k(\Lambda)}{ k!\,4^k},\quad k=1,\dots,N,\quad\quad c_0\neq 0,
\ee
where $\chi_j(\lambda)$ is the determinant of the $j\times j$  lower-right block \footnote{For example,
$$ 
\begin{aligned}
&
\chi_1(\lambda)= \mathcal{M}_{N+1,N+1}+\lambda,
\quad 
\chi_2(\lambda)=
\det\left( \begin{pmatrix}
\mathcal{M}_{NN} & \mathcal{M}_{N,N+1}
\\
\mathcal{M}_{N+1,N} & \mathcal{M}_{N+1,N+1}
\end{pmatrix}+\lambda\right),
\\
&
 \chi_3(\lambda)
=
\det\left(
 \begin{pmatrix}
\mathcal{M}_{N-1,N-1} & \mathcal{M}_{N-1,N}& 0 
\\
 \mathcal{M}_{N,N-1} & \mathcal{M}_{N,N}& \mathcal{M}_{N,N+1}
\\
0 & \mathcal{M}_{N+1,N} & \mathcal{M}_{N+1,N+1}
\end{pmatrix}+\lambda\right),\quad .\, .\, .\,\quad ,  \chi_{N+1}(\lambda)=\det(\mathcal{M}+\lambda).
\end{aligned}
$$
} of $\mathcal{M}(b,M,N)+\lambda I_{N+1}$.
They form  an eigenvector  
   of $\mathcal{M}$  relative to $-\Lambda$, i.e., 
  \be
\label{30aprile2024-5-bis}
\bigl(\mathcal{M}(b,M,N)+\Lambda\bigr)\boldsymbol{c}=0,\quad\quad\hbox{where }\boldsymbol{c}:= 
 \begin{pmatrix} c_{2N}
  \\
  c_{2N-2}
  \\
  \vdots
  \\
  c_2
  \\
  c_0
  \end{pmatrix}.
\ee
   The geometric multiplicity of $-\Lambda$ in the eigenvalue problem \eqref{30aprile2024-5-bis} is  one.   
\epr
 
\bre
\label{22aprile2026-2}
 {\rm 
In the above proposition, $Q_1(x,\Lambda)$ is completely determined by $c_0, M, N$ and the eigenvalue $\Lambda$.   The characteristic polynomial of $\mathcal{M}$ depends   on $M,N$ polynomially, but not on $c_0$. The coefficient $c_{2k}=c_{2k}(\Lambda)$ is a polynomially  of degree $k$ in $\Lambda$. }
\ere

\vskip 0.2 cm 
\begin{proof} 
Substitution of an expression $ 
y(x)= Q(x) e^{-\vartheta(x)}
$ 
 into \eqref{30aprile2024-2} yields the following ODE:
$$ 
E_\lambda(Q)=0
.
$$
The  operator $$
E_\lambda:=\frac{d^2}{dx^2} -2(x^3+bx)\frac{d}{dx}+\left(2Mx^2+\lambda-b-\frac{\gamma}{x^2}\right)
$$
 preserves the linear space over $\mathbb{C}$ generated by the elementary functions $x^{M-2k}$, $k\in\mathbb{N}$,  because
$$ 
E_\lambda(x^{M-2k})= x^{M-2k}\left(4kx^2+ \{A(k)+\lambda\}+ \frac{B(k)}{x^2}\right),\quad k\geq 0,
$$
where 
$$ 
A(k):= b(4k-2M-1) ,
\quad 
B(k):=( 2k-M)( 2k + 1-M) - \gamma.
$$
Substituting  
$$Q(x):=\sum_{k=0}^\infty c_{2k}x^{-2k} \cdot  x^M$$
 into $E_\lambda(Q)=0$, we obtain the recurrence relations
\be
\label{30aprile2024-3}
\left\{
\begin{aligned}
&4c_2+(A(0)+\lambda)c_0=0,
\\
\noalign{\medskip}
& 4(k+1)c_{2(k+1)}+(A(k)+\lambda)c_{2k}+B(k-1)c_{2(k-1)}=0, \quad\quad k\geq 0
\end{aligned}
\right.
\ee
represented by a semi-infinite tridiagonal matrix extending to the right and down:  \be
\label{6maggio2026-1-bis}
\overbrace{\begin{pmatrix}
A(0) & 4
\\
\noalign{\medskip}
B(0) & A(1) & 8 
\\
\quad\quad \ddots&\quad\quad\ddots&\quad\quad\ddots
\\
\mathrm{row}\, j\,\rightarrow&B(j-2) &A(j-1) &4 j 
\\
& \quad\quad\ddots&\quad\quad\ddots&\quad\quad\ddots
\\
&&&&
\end{pmatrix}}^{\mathcal{C}_\infty(b,M,N)}
\begin{pmatrix} c_0
  \\
  c_2
  \\
  c_4
  \\
  \vdots
  \\
  c_{2j}
  \\
  c_{2(j+1)}
  \\
  c_{2(j+2)}
  \\
  \vdots
  \end{pmatrix}=-\lambda\begin{pmatrix} c_0
  \\
  c_2
  \\
  c_4
  \\
  \vdots
  \\
  c_{2j}
  \\
  c_{2(j+1)}
  \\
  c_{2(j+2)}
  \\
  \vdots
  \end{pmatrix}
\ee
with entries
 $$
\mathcal{C}_\infty(b,M,N)_{jk}
:=A(j-1)\delta_{jk}+B(j-2)\delta_{j,k+1}+4j\,\delta_{j,k-1},
\quad\quad  j,k\geq 1.
$$
 Note that $c_0=0$ will imply that all $c_{2k}=0$ for all $k$. Assuming  that $c_{2N}\neq 0$  for some $N\geq 0$, the above recurrences have a solution with  $c_{2k}=0$ for all $k\geq N+1$ if and only if 
$$
 c_{2N}\, \cdot(\mathcal{C}_\infty(b,M,N))_{N+2,N+1} =0,
$$
and $\lambda$ is such that the kernel of the matrix 
$$ \mathcal{C}_{N+1}(b,M)+\lambda I_{N+1}
$$
is not trivial, where we have defined the truncation
$$\mathcal{C}_{N+1}(b,M):=\hbox{ $(N+1)\times(N+1)$ upper-left block of } \mathcal{C}_\infty(b,M,N).
$$ 
Since $c_{2N}\neq 0$ by assumption,  the necessary and sufficient condition is
 $$B(N)\equiv (\mathcal{C}_\infty(b,M,N))_{N+2,N+1}=0
 $$
 and
 $$
  \lambda=\Lambda \hbox{  eigenvalue of }-\mathcal{C}_{N+1}(b,M).
 $$
Since  $B(N)=0$ if and only if 
 $\gamma=(2N-M)( 2N + 1-M),
 $  we have proved conditions a) and b).\footnote{Note that  $E_\lambda$ preserves the  finite-dimensional linear space over $\mathbb{C}$ generated by $x^{M-2k}$, $k=0,1,\dots,N$, if and only if $B(N)=0$. 
Therefore,  $\gamma=(2N-M)( 2N + 1-M)$ is clearly a sufficient condition to have a non-trivial solution of $E_\Lambda(Q)=0$ of the form $ 
Q_1(x)= \sum_{k=0}^N c_{2k}x^{-2k} ~ x^M, \quad c_0\neq 0.
$}

\vskip 0.2 cm 
We take now $\gamma=(2N-M)( 2N + 1-M)
 $, so that  $B(k)$ becomes   
$$B(k)=2(N - k)( 2M-2N  - 1 - 2k),
$$  
and 
$$
\mathcal{C}_\infty(b,M,N)_{jk}= b(4j-2M-5)\delta_{jk}+2(N+2-j)(2M-2N-2j+3)\delta_{j,k+1}+4j\,\delta_{j,k-1}.
$$
We define\footnote{The reverted matrix of  a $n\times n $ matrix $\mathcal{A}=(\mathcal{A}_{jk})_{j,k=1}^n$, 
is the matrix with entries 
$ \bigl(
( \mathcal{A})_{n-j+1,n-k+1}
\bigr)_{j,k=1}^n
,$ 
obtained by reversing the order of both rows and columns.}
\be
\label{5giugno2026-5}
\mathcal{M}(b,M,N) := \hbox{ reverted matrix of } \mathcal{C}_{N+1}(b,M).\ee
We conclude   that there exists a  solution $$ 
Q_1(x)= \sum_{k=0}^N c_{2k}x^{-2k} ~ x^M, \quad c_0\neq 0,
$$
   if and only if $\boldsymbol{c}=(c_{2N},\dots,c_2,c_0)^{\mathrm{T}}$ is an eigenvector of $\mathcal{M}(b,M,N)$ and $-\lambda$ is equal to an eigenvalue $-\Lambda$.

\vskip 0.2 cm 
Since the recurrence relations \eqref{30aprile2024-3},  for $\Lambda$ a root of $ 
\det(\mathcal{M}+\lambda)=0
$, uniquely determine the ratio  $c_{2k}/c_0$, $k=1,\dots,N$., then  the eigenvectors of $\mathcal{M}(b,M,N)$  relative to an eigenvalue $-\Lambda$ span a 1-dimensional space.

\vskip 0.2 cm 
The explicit solution of the recurrence relation is as follows. The adjugate matrix $\mathrm{adj}(\mathcal{M}(b,M,N)+\lambda)$  satisfies 
\be
\label{3giugno2026-1}
(\mathcal{M}(b,M,N)+\lambda  )\cdot\mathrm{adj}(\mathcal{M}(b,M,N)+\lambda ) =\chi_{N+1}(\lambda) I_{N+1},
\ee
where 
$$ \chi_{N+1}(\lambda):= \det(\mathcal{M}(b,M,N)+\lambda ).
$$
 Hence, if $-\lambda=-\Lambda$ is an eigenvalue of $\mathcal{M}$, we receive
$$
(\mathcal{M}(b,M,N)+\Lambda  )\cdot\mathrm{adj}(\mathcal{M}(b,M,N)+\Lambda ) =0.
$$
It follows  that
$$ 
\boldsymbol{v}_j(\Lambda):=\mathrm{adj}(\mathcal{M}(b,M,N)+\Lambda )\,\boldsymbol{e}_j
,
$$
is an eigenvector, if it is not zero (here $\boldsymbol{e}_1,\cdots,\boldsymbol{e}_{N+1}$ is the standard basis of columns in $\mathbb{C}^{N+1}$).

\vskip 0.2 cm 
If $-\Lambda$ {\it has geometric multiplicity equal to one}, as in our case, the following properties hold \cite{Gant}:\footnote{Given an $n\times n$ matrix $A$, from 
$ \mathrm{adj}(A)\,A=A\,\mathrm{adj}(A)=\det(A)\,I$ it follows that 
\begin{itemize}
\item $ \mathrm{rank} A=n$  $\Leftrightarrow$ $ \mathrm{rank} (\mathrm{adj}A)=n$;
\item $ \mathrm{rank} A=n-1$  $\Leftrightarrow$ $ \mathrm{rank} (\mathrm{adj}A)=1$;
\item $ \mathrm{rank} A\leq n-2$  $\Leftrightarrow$ $  \mathrm{adj}A=0$.
\end{itemize}
} 
\begin{itemize} 
\item
 $ 
\mathrm{adj}(\mathcal{M}+\Lambda  )\neq 0$, and $ \mathrm{rank}( \mathrm{adj} (\mathcal{M}+\Lambda))=1$;

\item
          $\exists\, j\in\{1,\dots, N+1\}$ such that $\boldsymbol{v}_j(\Lambda)\neq 0$, so that it is an eigenvector.
\end{itemize}
Note that our tridiagonal matrix has structure  
\be
\label{22aprile2026-1}
\mathcal{M}(b,M,N)=\begin{pmatrix}
* & * & 
\\
\beta_N & * & * 
\\
& \beta_{N-1}& * & * 
\\
\\
&&\ddots&
\ddots&\ddots&
\\
\\
&&&
\beta_2&*&*
\\
&&&&\beta_1&*
\end{pmatrix},\quad\quad \beta_j:=4 j\neq 0
\ee
  so it is straightforward to compute 
\be
\label{16giugno2026-5}
\boldsymbol{v}_1(\lambda)=\mathrm{adj}\bigl(\mathcal{M}(b,M,N)+\lambda \bigr)\,\boldsymbol{e}_1 
= 
\begin{pmatrix}   \chi_N(\lambda) 
  \\
  \noalign{\medskip}
  -\beta_N \,  \chi_{N-1}(\lambda) 
  \\
  \noalign{\medskip}
  \vdots
  \\
  \noalign{\medskip}
(-1)^{N-2} \beta_3\cdots\beta_N \, \chi_2(\lambda)  \\
  \noalign{\medskip}
(-1)^{N-1} \beta_2\beta_3\cdots\beta_N \, \chi_1(\lambda) 
  \\
  \noalign{\medskip}
 (-1)^N\beta_1\beta_2\beta_3\cdots\beta_N
  \end{pmatrix}
\ee
This is the eigenvector we are looking for, because  $ \boldsymbol{v}_1(\Lambda)\neq 0$, being  $\beta_1\cdots\beta_N\neq 0$. Equivalently, we can choose the eigenvector 
\be
\label{16giugno2026-12}
\boldsymbol{c}
:= c_0\, \frac{(-1)^N  \,\boldsymbol{v}_1}{ \beta_1\beta_2\cdots\beta_N},
\ee 
which yields \eqref{27aprile2026-2}.\end{proof}

\bre{\rm 
The eigenfunctions 
 $
 y_1(x,\Lambda)$  of   Proposition \ref{13aprile2023-4}, corresponding to a $\Lambda$ in the algebraic spectrum,  form a 1-parameter space, the parameter being $c_0$. They  satisfy  the boundary conditions 
\be
\label{16giugno2026-10}
y(x)\to 0 \hbox{ for $ x\to +\infty e^{i\pi\ell/2}$, with $\ell\in\mathbb{Z}$}.
\ee
By analysing the Stokes phenomenon of  \eqref{30aprile2024-2}, one can actually prove the following theorem: 
{\it 
Equation \eqref{30aprile2024-2} with boundary conditions \eqref{16giugno2026-10} has non trivial solution if and only if 
$ \gamma=(2N-M+1)(2N-M)$ for some $N\in\mathbb{N},
$
 and $\lambda$ is equal to an element $\Lambda$ in the   algebraic spectrum. 
In this case, the solution is 
$
y(x)=y_1(x,\Lambda)$ 
as in Proposition \ref{13aprile2023-4}.}  
We will not delve into this problem here.
} 
\ere

\subsubsection{Asymptotic solutions $y_1^{(\nu)}(x,\lambda)$ with negative exponential, $\lambda\in\mathbb{C}$}

Whatever is  $\gamma\in\mathbb{C}$, it is a well known standard computation to show that the ODE \eqref{30aprile2024-2}  admits a formal solution at $x=\infty$ with  structure
$$ 
y_1^{(F)}(x,\lambda)=\sum_{k=0}^\infty c_{2k}(\lambda) x^{-2k} \,x^M \,e^{-\vartheta(x)},
$$ 
where $c_{2k}(\lambda)$ is a polynomial of degree $k$ in $\lambda$, determined by 
the  recurrence relations \eqref{30aprile2024-3}.
 By \eqref{6maggio2026-1-bis}, we see that
\be
\label{6maggio2026-1}
c_{2k}(\lambda)=c_0 \, \frac{(-1)^k\,\chi_k(\lambda)}{k!\,4^k},\quad k\in \mathbb{N},
\ee
where $\chi_k(\lambda)$ is the determinant of the upper-left block of dimension $k\times k$ of $\mathcal{C}_\infty$. For $k=0,1,...,N+1$ it is  exactly the same $\chi_k(\lambda)$ of Proposition \ref{13aprile2023-4}.

\begin{figure}
\centerline{\includegraphics[width=0.6\textwidth]{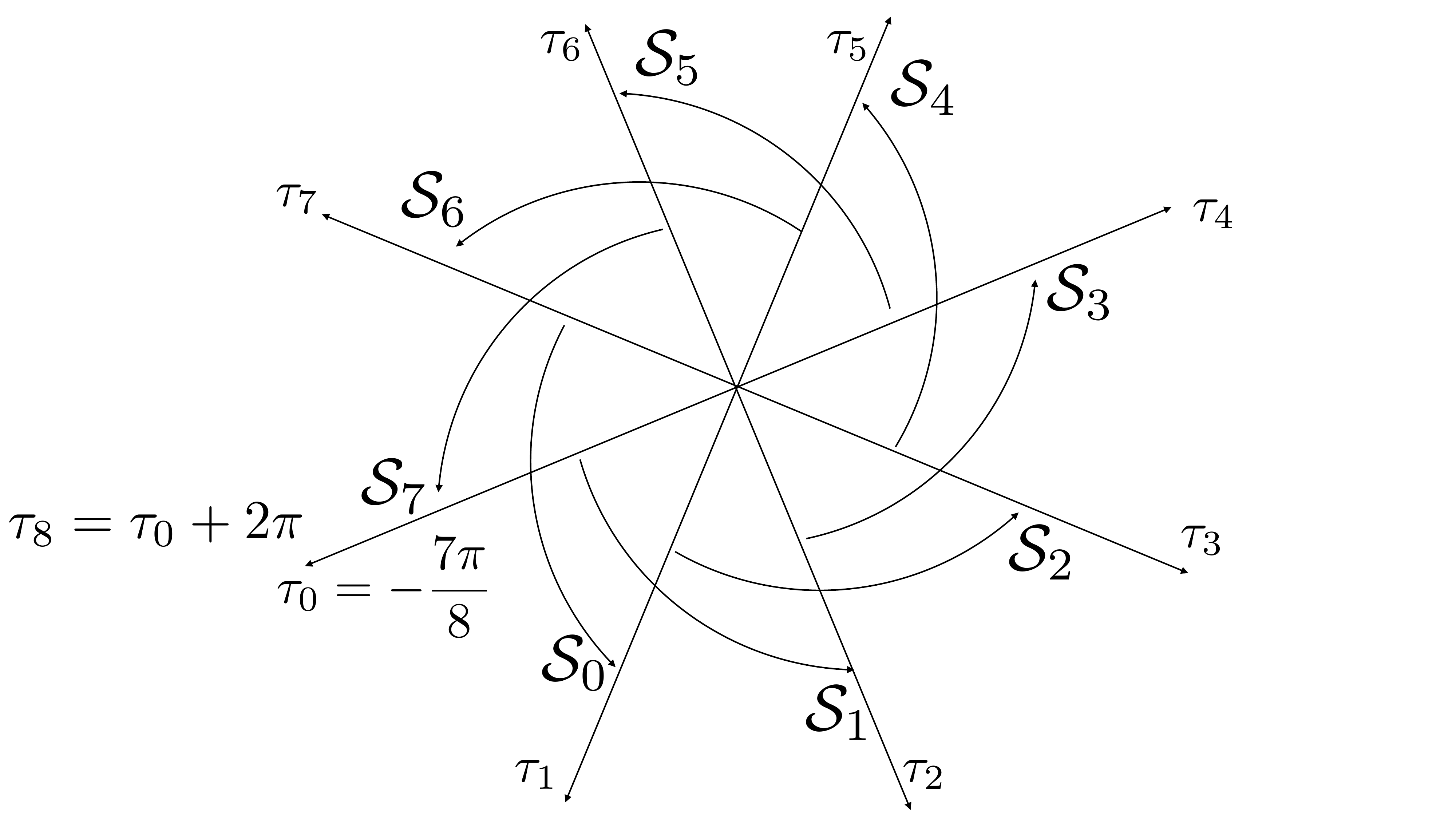}}
\caption{Sectors. The open sector $\mathcal{S}_\nu=S(\tau_{\nu-1},\tau_{\nu+1})$  contains only the Stokes ray $\arg x=\tau_\nu$.}
\label{13aprile2024-5}
\end{figure}

\vskip 0.2 cm 
The  asymptotic theory of ODEs \cite{Sh4, Sh2, Wasow,CGM2023} prescribes the existence of a unique  actual solution asymptotic to the formal one, as follows. In the universal covering $\mathcal{R}$ of  $\mathbb{C}\backslash\{0\}$,   the {\it Stokes rays} are defined by 
\be
\label{16giugno2026-8}
\arg x = \tau_\nu,\quad\quad \hbox{ where }\quad \tau_\nu:=\frac{2\nu-7}{8}\pi,\quad \nu\in\mathbb{Z}.
\ee
Consider the sectors (see Figure \ref{13aprile2024-5})
$$
\mathcal{S}_\nu:=S(\tau_{\nu-1},\tau_{\nu+1})=\{x\in \mathcal{R}~|~ \tau_{\nu-1}<\arg x<\tau_{\nu+1}\}.
$$
For each $\nu\in\mathbb{Z}$ there is a unique actual solution $y_1^{(\nu)}(x,\lambda)$ of  \eqref{30aprile2024-2}, holomorphic of $(x,\lambda)\in \mathcal{R}\times \mathbb{C}$, characterized by the asymptotic behaviour
$$ 
y_1^{(\nu)}(x,\lambda)\sim y_1^{(F)}(x,\lambda),\quad x\to\infty \hbox{ in } \mathcal{S}_{2\nu-1}\cup\mathcal{S}_{2\nu}.
$$

\subsubsection{Proposition \ref{13aprile2023-4} revised}

The meaning of Proposition \ref{13aprile2023-4} is that, for any $\nu\in\mathbb{Z}$, the solution $y_1^{(\nu)}(x,\lambda) $ becomes a quasi-polynomial solution $y_1(x,\Lambda)$  if and only if $\gamma=(2N-M+1)(2N-M)$ and $\lambda$ becomes  equal to an eigenvalue $\Lambda$ of $-\mathcal{M}(b,M,N)$, or equivalently of $-\mathcal{C}_{N+1}(b,M)$. 
This follows also from the structure of $\mathcal{C}_\infty$.
 When $B(N)=0$, i.e., when  $\gamma=(2N-M+1)(2N-M)$, the structure is 
$$ 
\mathcal{C}_\infty=
\left(
\begin{array}{c|c}
\mathcal{C}_{N+1} &{O}_{N+1}  \\
&
\\
\hline
&
\\
\boldsymbol{0} & H
\end{array}
\right.   \,\,,\quad\quad H= \begin{array}{|ccccc} \hline * & * & 0& 0 &\cdots
\\
* & * & * & 0 &\cdots
\\
0 & \ddots&\ddots& \ddots &
\end{array}
$$
where $H$ is a semi-infinite tridiagonal matrix, ${O}_{N+1}$ has $N+1$ rows and is semi-infinite to the right, with all  entries equal to zero, except for the entry  $(\mathcal{C}_\infty)_{N+1,N+2}=4(N+1)$:
$$ 
{O}_{N+1}=\left(\begin{array}{ccc}
0 & 0 & \cdots
\\
\vdots & \vdots & 
\\
0 & 0 & \cdots\\
4(N+1)& 0 & \cdots
\end{array}\right. 
.
$$
Therefore, for $k\geq N+1$ the determinanant of the $k\times k$ submatrices of $\mathcal{C}_\infty+\lambda$ factorizes as 
\be
\label{3giugno2026-5}
\chi_k(\lambda)=\chi_{N+1}(\lambda) \, \mathfrak{X}_{k-N-1}(\lambda),\quad\hbox{  $k\geq N+1$},
\ee
where $\mathfrak{X}_0:=1$, while $\mathfrak{X}_{k-N-1}$ is a polynomial of degree $k-N-1$ given by  the determinant of the upper-left block of dimension $k-N-1$ of $H+\lambda$. It follows that  if $\Lambda$ is an eigenvalue, i.e.,
$$\left. \begin{aligned}
& \det(\mathcal{M}(b,M,N)+\Lambda )
\\
\noalign{\medskip}
&\det\bigl(\mathcal{C}_{N+1}(b,M)+\Lambda \bigr)
\end{aligned}
\right\}
=\chi_{N+1}(\Lambda)=0, $$
then
$$
\chi_k(\Lambda)=\underbrace{\chi_{N+1}(\Lambda)}_{=0} \, \mathfrak{X}_{k-N-1}(\Lambda)=0 \quad\Longrightarrow \quad c_{2k}(\Lambda)=0 \quad\hbox{ for $k\geq N+1$}.
$$

\subsection{Quasi-polynomial Solutions with positive exponential}

The analogue of Proposition \ref{13aprile2023-4} holds for quasi-polynomial solutions with positive exponential.  
\bpr
\label{1febbraio2026-1}
The eigenvalue problem   \eqref{30aprile2024-2}  has a solution $(\Lambda, y_2(x,\Lambda))$ with 
$$
y_2(x,\Lambda):=Q_2(x,\Lambda) e^{\vartheta(x)},\quad \quad Q_2(x,\Lambda)=\Bigl(\sum_{k=0}^N d_{2k}(\Lambda) \,x^{-2k}\Bigr)\, x^{-M-3},\quad N\in\mathbb{N}
$$
if and only if the following conditions hold. \begin{itemize}

\item[$a)$] $\gamma=(2N+M+4)(2N+M+3)$ for some $N\in\mathbb{N}$; 

\item[$b)$] $\lambda=\Lambda$, a root of the characteristic polynomial
\be
\label{18febbraio2026-6}
\det\Bigl(\pm i\mathcal{M}(\pm ib,-M-3,N)+\lambda\Bigr)=0 .
\ee
where the functional dependence of a matrix $\mathcal{M}(b,M,N)$ is defined in \eqref{28maggio2024-1}. 
\end{itemize}
  The coefficients are
  \be
\label{5maggio2026-7}
d_{2k}(\Lambda)=d_0\,\frac{\widetilde{\chi}_k(\Lambda)}{k!\,4^k},\quad k=1,\dots,N,\quad d_0\neq 0, 
\ee
where 
$\widetilde{\chi}_j(\lambda) $ is the determinant of the lower-right $j\times j$ block of $\bigl(\pm i{\mathcal{M}(\pm i b, -M-3,N})+\lambda \bigr)$, the signs $\pm$ giving the same result. They form  an eigenvector  $\boldsymbol{d}$ 
     relative to $-\Lambda$ for the eigenvalue problem 
     \be
  \label{5maggio2026-1}
\left\{\mathfrak{I}_{\pm}^{-1}  \Bigl(
\pm i\,\mathcal{M}(\pm i b,\,-M-3,\, N)\Bigr)\, \mathfrak{I}_{\pm}\right\}\,\boldsymbol{d} =
-\Lambda \boldsymbol{d},\quad\quad \boldsymbol{d}=\begin{pmatrix} 
{d}_{2N}
  \\
  \vdots
  \\
{d}_2
  \\
{d}_0
  \end{pmatrix}
  \ee
  where $\mathfrak{I}_{\pm}:=\mathrm{diag}\bigl((\pm i)^N,\,(\pm i)^{N-1},\,\dots,\,\pm i, \, 1\bigr)$. The matrix in \eqref{5maggio2026-1} does not depend on the sign $\pm$.
\epr

\begin{proof}
Equation \eqref{30aprile2024-2} admits the symmetry
\be
\label{23giugno2026-1}
\begin{aligned}
& \widetilde{x}= e^{\pm i\pi/4} x,
& \widetilde{b}= \pm i b,
\\
& \widetilde{M}= -3-M,
&\widetilde{\lambda}= \mp i\,\lambda,
\end{aligned}\quad\quad \hbox{$\gamma$  unchanged}.
\ee
Indeed,  by \eqref{23giugno2026-1}, equation \eqref{30aprile2024-2} becomes
$$
-\frac{d^2y}{dx^2}+\left(\widetilde{x}\,^6+2\widetilde{b}\,\widetilde{x}\,^4+(\widetilde{b}^2-2\widetilde{M}-3)\widetilde{x}\,^2+\frac{\gamma}{\widetilde{x}\,^2}\right)y=\widetilde{\lambda}y
$$
By Proposition \ref{13aprile2023-4}, the above problem has quasi-polynomial solution with negative exponential   
$$ 
y_1(\widetilde{x},\widetilde{\Lambda})=\sum_{k=0}^N \widetilde{c}_{2k}(\widetilde{\Lambda}) \,\widetilde{x}^{-2k} \, \widetilde{x}^{\widetilde{M}}\,e^{-\vartheta(\widetilde{x})},
$$
if and only if $\gamma=(2N-\widetilde{M}+1)(2N-\widetilde{M})$ and $\widetilde{\lambda}=\mp i \lambda$  is equal to a root $\widetilde{\Lambda}$ of the characteristic polynomial
 $$
\det(\mathcal{M}\bigl(\,  \pm i b,\, -M-3,\,N\bigr)+\widetilde{\lambda})=0.
$$ 
The coefficients $\widetilde{c}_{2k}$ form an eigenvector  $\widetilde{\boldsymbol{c}}$  relative to $-
 \widetilde{\Lambda}$, i.e.,
$$
\Bigl(\mathcal{M}\bigl(\,  \pm i b,\, -M-3,\,N\bigr)+\widetilde{\Lambda}\Bigr)\widetilde{\boldsymbol{c}}
= 0,\quad \quad \widetilde{\boldsymbol{c}}=\begin{pmatrix} \widetilde{c}_{2N}
  \\
  \widetilde{c}_{2N-2}
  \\
  \vdots
  \\
  \widetilde{c}_0
  \end{pmatrix},\quad \widetilde{c}_0\neq 0.
$$
Due to \eqref{23giugno2026-1}, we have proved conditions a) and b), with the desired quasi-polynomial solution 
$$
y_2(x,\Lambda)=y_1(\widetilde{x},\widetilde{\Lambda}),
$$
with 
$ 
  d_{2k}= e^{\mp i {\pi}(M+3)/4}\, \exp\left\{\mp\, \frac{i \pi\,k}{2}\right\}\,\widetilde{c}_{2k}.
  $
The factor $e^{\mp i \frac{\pi}{4}(M+3)}$ is inessential, being $\alpha\, y_2(x,\Lambda)$ a quasi-polynomial solution for any $\alpha\in \mathbb{C}\backslash\{0\}$. Dropping this factor,  we receive  $ 
 d_0= \widetilde{c}_0\neq 0$,  $ d_{2k}= \exp\left\{\mp\, \frac{i \pi\,k}{2}\right\}\,\widetilde{c}_{2k},
$ 
and  
$$ \Bigl(
\pm i\,\mathcal{M}(\pm i b,\,-M-3,\, N)\,+\Lambda
\Bigr)\begin{pmatrix} 
 e^{\pm i \pi N/2} d_{2N}
    \\
  \vdots
  \\
  e^{\pm i \pi k/2} d_{2k}
  \\
  \vdots
  \\
  {d}_0
  \end{pmatrix}
=0.
$$
The above equation is exactly \eqref{5maggio2026-1}. The fact that  $ \mathfrak{I}_{\pm}^{-1}  \Bigl(
\pm i\,\mathcal{M}(\pm i b,\,-M-3,\, N)\Bigr)\, \mathfrak{I}_{\pm}$ is independent of the choice of sign  $\pm$ will follow from \eqref{5giugno2026-7} below.

The explicit form of an eigenvector relative to an eigenvalue $-\Lambda$ of $\pm i\,\mathcal{M}(\pm i b,\,-M-3,\, N)$ (recall that the geometric multiplicity is 1) is 
$${\small
\begin{aligned}
\widetilde{\boldsymbol{v}}_1(\Lambda)&:=\mathrm{adj}\Bigl(\pm i\,\mathcal{M}(\pm i b,\,-M-3,\, N)+\Lambda\Bigr)\,\boldsymbol{e}_1
=\begin{pmatrix} 
 {}^{\pm}\widehat{\chi}_N(\Lambda)
    \\
    \noalign{\medskip}
    -(\pm i\beta_N) \,\cdot\, {}^{\pm}\widehat{\chi}_{N-1}(\Lambda)
    \\
    \noalign{\medskip}
   (\pm i \beta_{N-1})(\pm i\beta_N)\,\cdot\,  {}^{\pm}\widehat{\chi}_{N-2}(\Lambda)
    \\
    \noalign{\medskip}
  \vdots
  \\
   \noalign{\medskip}
 (-1)^{N-1}(\pm i\beta_2)\cdots (\pm i \beta_{N-1})(\pm i\beta_N)\,\cdot\,  {}^{\pm}\widehat{\chi}_{1}(\Lambda)
  \\
   \noalign{\medskip}
(-1)^N (\pm i \beta_1)(\pm i \beta_2)\cdots (\pm i \beta_{N-1})(\pm i\beta_N) \quad\quad
  \end{pmatrix}
\neq 0,
\end{aligned}
}$$
where 
${}^{\pm}\widehat{\chi}_j(\lambda) $ is the determinant of the lower-right $j\times j$ block of $\bigl(\pm i{\mathcal{M}(\pm i b, -M-3,N})+\lambda \bigr)$. 
 We can  take the renormalized eigenvector
\be
\label{21aprile2026-2}
\widetilde{\boldsymbol{c}}=\dfrac{(\pm i)^N \,\widetilde{c}_0\,\widetilde{\boldsymbol{v}}_1(\Lambda)}{\beta_1\cdots\beta_N}
\ee
% The eigenvector of $\widetilde{\mathcal{M}}_{\pm}(b,M,N)$ relative to $\widetilde{\Lambda}_0$ is then 
% $$
% \widetilde{\boldsymbol{c}}=\begin{pmatrix} \widetilde{c}_{2N}
%   \\
%   \widetilde{c}_{2N-2}
%   \\
%   \vdots
%   \\
%   \widetilde{c}_2
%   \\
 %  \widetilde{c}_0
 %  \end{pmatrix}
% = \widetilde{c}_0\cdot 
% \begin{pmatrix}   (-1)^N\dfrac{  \widetilde{\chi}_N(\widetilde{\Lambda}_0)}{\beta_1\beta_2\beta_3\cdots \beta_N}
%   \\
 %  \noalign{\medskip}
 %  \vdots
 %  \\
 %  \noalign{\medskip}
% -\dfrac{  \widetilde{\chi}_3(\widetilde{\Lambda}_0)}{\beta_1\beta_2\beta_3}
 %  \\
 %  \noalign{\medskip}
 % \dfrac{ \widetilde{\chi}_2(\widetilde{\Lambda}_0)}{\beta_1\beta_2}
 %  \\
%   \noalign{\medskip}
%  -\dfrac{\widetilde{\chi}_1(\widetilde{\Lambda}_0)}{\beta_1}
%   \\
 %  \noalign{\medskip}
 %  1
%   \end{pmatrix}
% $$
% where $$(\mathcal{\widetilde{M}}_\pm)_{N-j+2,N-j+1}\equiv\mathcal{M}_{N-j+2,N-j+1}=\beta_j=4j$$
% and 
% $$
% \widetilde{\chi}_j(\widetilde{\Lambda}_0) = \hbox{  determinant of  lower right $j\times j$ submatrix of $\widetilde{\mathcal{M}}_\pm+\widetilde{\Lambda}_0$}.
% $$ 
 It follows 
 that (with $d_0=\widetilde{c}_0$)
$$
\boldsymbol{d}=\mathrm{diag}\Bigl((\mp i)^N,\, (\mp i)^{N-1},\,\dots,\,\mp i,\,1\Bigr) \,\widetilde{\boldsymbol{c}}.
$$
This is exactly \eqref{5maggio2026-7}, where we have defined
$$
\widetilde{\chi}_j(\lambda):={}^{\pm}\widehat{\chi}_j(\lambda),
$$ 
This is a well posed definition,  because 
 ${}^{+}\widehat{\chi}_j(\lambda) $  and ${}^{-}\widehat{\chi}_j(\lambda) $ are equal. Their equality  will follow from \eqref{5giugno2026-7} below.

\end{proof}

\subsubsection{Asymptotic solutions $y_2^{(\nu)}(x,\lambda)$ with positive exponential and Proposition \ref{1febbraio2026-1} revised}

Proposition \ref{1febbraio2026-1} can also be proved directly as for  Proposition \ref{13aprile2023-4}, substituting  into the equation the expression $y(x)=Q(x) \exp\{\vartheta(x)\}$. This gives $\widetilde{E}_\lambda (Q)=0$, where 
$$ 
\widetilde{E}_\lambda:=
\frac{d^2}{dx^2} +2(x^3+bx)\frac{d}{dx}+2(M+3)x^2+\Lambda+b-\frac{\gamma}{x^2}
.
$$
Then,
$$ 
\widetilde{E}_\lambda(x^{-M-3-2k})=x^{-M-3-2k}\left(-4kx^2+\bigl\{\widetilde{A}(k)+\lambda\bigr\} +\frac{\widetilde{B}(k)}{x^2}\right),\quad k\geq 0,
$$
with 
$$ 
\widetilde{A}(k):= -b(2M+4k+5),
\quad 
\widetilde{B}(k):= (M+2k+3)(M+2k+4)-\gamma.
$$
Substitution of 
$$Q(x)=\sum_{k=0}^\infty d_{2k}x^{-2k}\, x^{-M-3} 
$$
into $\widetilde{E}_\lambda (Q)=0$ gives the recurrence relations 
$$ \left\{
\begin{aligned}
&
-4 d_2+(\widetilde{A}(0)+\lambda) d_0 =0,
\\
&
-4(k+1)d_{2(k+1)}+(\widetilde{A}(k)+\lambda) d_{2k} +\widetilde{B}(k-1)d_{2(k-1)}=0.
\end{aligned}
\right.
$$
They can be represented as the eigenvalue problem of a semi-infinite matrix $\mathcal{D}_\infty(b,M,N)$  with entries 
$$ 
\mathcal{D}_\infty(b,M,N)_{jk}:=\widetilde{A}(j-1)\, \delta_{jk} +\widetilde{B}(j-2)\,\delta_{j,k+1}-4j \,\delta_{j,k-1},\quad\quad j,k\geq 1,
$$
that is 
\be
\label{6maggio2026-3}
\overbrace{\begin{pmatrix}
\widetilde{A}(0) & -4
\\
\noalign{\medskip}
\widetilde{B}(0) & \widetilde{A}(1) & -8 
\\
\quad\quad \ddots&\quad\quad\ddots&\quad\quad\ddots
\\
\mathrm{row}\, j\,\rightarrow&\widetilde{B}(j-2) &\widetilde{A}(j-1) &-4 j 
\\
& \quad\quad\ddots&\quad\quad\ddots&\quad\quad\ddots
\\
&&&&
\end{pmatrix} }^{\mathcal{D}_\infty(b,M,N)}
\begin{pmatrix} d_0
  \\
  d_2
  \\
  d_4
  \\
  \vdots
  \\
  d_{2j}
  \\
  d_{2(j+1)}
  \\
  d_{2(j+2)}
  \\
  \vdots
  \end{pmatrix}=-\lambda\begin{pmatrix} 
  d_0
  \\
  d_2
  \\
  d_4
  \\
  \vdots
  \\
  d_{2j}
  \\
  d_{2(j+1)}
  \\
  d_{2(j+2)}
  \\
  \vdots
  \end{pmatrix}
\ee
The above has solution 
\be
\label{6maggio2026-3-bis}
d_{2k}(\lambda)=d_0\, \frac{\widetilde{\chi}_k(\lambda)}{4^k\,k!},\quad k\in\mathbb{N},
\ee
where 
$$\widetilde{\chi}_k(\lambda):= \det(k\times k  \hbox{  upper-left block of }\mathcal{D}_\infty+\lambda I_k).
$$
We have $$\widetilde{B}(N)=0
\quad
\Longleftrightarrow\quad  \gamma=(2N+M+4)(2N+M+3). $$ 
In this case 
$
\widetilde{B}(k)= -2(N - k)(2N + 2M + 7 + 2k),
$
and
$$
\mathcal{D}_\infty=
\left(
\begin{array}{c|c}
\mathcal{D}_{N+1} & \widetilde{O}_{N+1} 
\\
&
\\
\hline
&
\\
\boldsymbol{0} & \widetilde{H}
\end{array}
\right.,\quad\quad \quad 
 \widetilde{H}=\begin{array}{|ccccc} \hline * & * & 0& 0 &\cdots
\\
* & * & * & 0 &\cdots
\\
0 & \ddots&\ddots& \ddots &
\end{array},
$$
where
$$\mathcal{D}_{N+1}(b,M):=\hbox{ $(N+1)\times(N+1)$ upper-left submatrix of } \mathcal{D}_\infty(b,M,N),
$$ 
 $\widetilde{H}$ is a semi-infinite tridiagonal matrix
and  $\widetilde{O}_{N+1}$ has $N+1$ rows and is semi-infinite to the right, with all zeros entries, except for  $(\mathrm{inv}\mathcal{\widetilde{\mathcal{M}}}_\infty)_{N+1,N+2}=-4(N+1)$:
$$
\widetilde{O}_{N+1}=\left(\begin{array}{ccc}
0 & 0 & \cdots
\\
\vdots & \vdots & 
\\
0 & 0 & \cdots\\
-4(N+1)& 0 & \cdots
\end{array}\right.  \,\,\,.
$$
Therefore,  if  $k\geq N+1$ the determinant $\widetilde{\chi}_k(\lambda)$ factorizes as 
\be
\label{3giugno2026-6}
\widetilde{\chi}_k(\lambda)=\widetilde{\chi}_{N+1}(\lambda) \, \widetilde{\mathfrak{X}}_{k-N-1}(\lambda), \quad \quad \hbox{  $k\geq N+1$},
\ee
where  $\widetilde{\mathfrak{X}}_{k-N_2-1}$  is is a polynomial of degree $k-N-1$ given by the determinant of the 
upper-left block of dimension $k-N-2$ of $\widetilde{H}+\lambda$.
\vskip 0.15 cm 

We conclude that the problem admits a finite  solution $d_0,d_2,\dots,d_{2N}$ and  $d_{2k}=0$ for al $k\geq N+1$  if and only if 
$ \gamma=(2N+M+4)(2N+M+3) $,  
and $\lambda$ is equal to an eigenvalue $\Lambda$ of $-\mathcal{D}_{N+1}(b,M)$, because in this case 
$$
\widetilde{\chi}_k(\Lambda)=\underbrace{\widetilde{\chi}_{N+1}(\Lambda)}_{=0} \, \widetilde{\mathfrak{X}}_{k-N-1}(\Lambda)=0 \quad\Longrightarrow \quad d_{2k}(\Lambda)=0 \quad\hbox{ for $k\geq N+1$}.
$$

The eigenvalue problem for $\mathcal{D}_{N+1}(b,M)$ is exactly the problem \eqref{5maggio2026-1}, by construction, and one can also check  that 
\be
\label{5giugno2026-7}
\hbox{reverted matrix of } \mathfrak{I}_{\pm}^{-1}  \bigl(
\pm i\,\mathcal{M}(\pm i b,\,-M-3,\, N)
\bigr) \, \mathfrak{I}_{\pm}
=\mathcal{D}_{N+1}(b,M).
\ee

When $\lambda$ is not a root of $\widetilde{\chi}_{N+1}(\lambda)$, then \eqref{30aprile2024-2} has a {\it formal solution} 
$$ 
y_2^{(F)}(x,\lambda)=\sum_{k=0}^\infty d_{2k}(\lambda)x^{-2k} \,x^{-M-3} e^{\vartheta(x)}.
$$ 
 To it,  for any $\nu\in\mathbb{Z}$, there corresponds a {\it unique actual solution} $y_2^{(\nu)}(x,\lambda)$, holomorphic of $(x,\lambda)\in \mathcal{R}\times\mathbb{C}$, characterized by the asymptotic behaviour 
$$
y_2^{(\nu)}(x,\lambda)\sim y_2^{(F)}(x,\lambda),\quad\quad x\to \infty \hbox{ in }\mathcal{S}_{2\nu}\cup\mathcal{S}_{2\nu+1}.
$$

In conclusion, the meaning of Proposition \ref{1febbraio2026-1} is that, for any $\nu\in\mathbb{Z}$, the solution $y_2^{(\nu)}(x,\lambda) $ becomes a quasi-polynomial solution $y_2(x,\Lambda)$  if and only if $\gamma=(2N+M+4)(2N+M+3)$ and $\lambda$ becomes  equal to an eigenvalue $\Lambda$ of $\mp i\,\mathcal{M}(\pm i b,\,-M-3,\, N)$, or equivalently of $-\mathcal{D}_{N+1}(b,M)$. 

\section{Eigenvalues with algebraic multiplicity greater than one}

We now address the main problem of this paper, namely the characterization of those values of $b$  for which an eigenvalue in the spectrum has algebraic multiplicity greater than one. Such a characterization will be obtained in terms of a resultant and the existence of solutions to a nonhomogeneous differential equation.

 Consider the case of Proposition \ref{13aprile2023-4}, with $\gamma=(2N-M+1)(2N-M)$.   The algebraic multiplicity of an eigenvalue $-\Lambda$ of $\mathcal{M}(b,M,N)$ is greater than one  if and only if  $b$ is a  root of the polynomial of degree $N(N+1)$\footnote{The degree $N(N+1)$ follows from the degree $N+1$ of the polynomials in  $\lambda$ given by $\det(\mathcal{M}(b)+\lambda)$,  and the degree $N$ of  $\frac{\partial }{\partial \lambda}\det(\mathcal{M}(b)+\lambda)$.}
\be
\label{14giugno2025-5}
{\rm res}_\lambda\left(\det(\mathcal{M}(b,M,N)+\lambda),~ \frac{\partial }{\partial \lambda}\det(\mathcal{M}(b,M,N)+\lambda)\right)=0.
\ee

Note that the algebraic multiplicity can be strictly greater than 2. For example, in case $N=2$, $M=3$ and $b=0$ we have 
$$\mathcal{M}=
\begin{pmatrix}
0 & -2 & 0 
\\
8 & 0 & 4
\\
0 & 4 & 0
\end{pmatrix},
\quad \det (\mathcal{M}+\lambda)=\lambda^3,\quad \hbox{ Jordan form }
J=
\begin{pmatrix}
0 & 1 & 0 
\\
0 & 0 & 1
\\
0 & 0 & 0
\end{pmatrix}.
$$ 

 %%%%%%%%
\subsection{Symmetries by reflection}
\label{21febbraio2026-3}
Observing for example  Figure \ref{21marzo2026-1}, it is evident that the roots of the resultant form a set invariant by reflection with respect to both the horizontal and vertical axes. This can be proved in a simple way. 

\bpr
\label{21febbraio2026-3-PRO}
Let $N\in\mathbb{N}$ and $M\in \mathbb{C}$. 
  The $N(N+1)$ roots $b$ (counting multiplicity) of either  \eqref{14giugno2025-5} or 
\eqref{14giugno2025-5-BIS} 
   form a set  invariant by a rotation of 180 degrees in the $b$-plane. 

\epr
\label{1febbraio2026-3}
\begin{proof}
It sufficies to observe that the resultants  depend on $b^2$ only.
\end{proof}

\bcr
Let $N\in\mathbb{N}$ and $M\in \mathbb{R}$. 
  The $N(N+1)$ roots $b$ (counting multiplicity) of either  \eqref{14giugno2025-5} or 
\eqref{14giugno2025-5-BIS}    form a set  invariant by  reflection with respect to both the horizontal and vertical axes.  In other words,  if $b$ is a root, also $\pm b$ and $\pm \overline{b}$ are roots. 

\ecr

\begin{proof}
It suffices to observe that if $M$ is real, then the coefficients of the resultant polynomial are real. We conclude using Proposition \ref{21febbraio2026-3-PRO}
\end{proof}
%%%%%%%

\subsection{Characterization in terms of quasi-polynomial solutions of the non homogeneous equation}

 Though we will not directly use the results of this subsection in the remainder of the paper, for the sake of completeness we provide a characterization of eigenvalues with algebraic multiplicity greater than one in terms of quasi-polynomial solutions to the nonhomogeneous equation. An alternative proof of the two lemmas below will be given in Appendix 2. 

\ble 
\label{3giugno2026-2}
Let $M\in \mathbb{C}$, $N\in\mathbb{N}$ and $\gamma=(2N-M+1)(2N-M)$ as in Proposition \ref{13aprile2023-4}, and let  $y_1(x;\Lambda)$ be the quasi-polynomilal solution of \eqref{30aprile2024-2} corresponding to and eigenvalue $-\Lambda$ of $\mathcal{M}(b,M,N)$. Then, $\Lambda$  has algebraic multiplicity greater than one  if and only if the non-homogeneus equation 
   \be
   \label{16giugno2026-1}
  \frac{d^2 w}{dx^2}+\left(\Lambda-\left(x^6+2bx^4+(b^2-2M-3)x^2+\frac{\gamma}{x^2}\right)\right)w= y_1(x;\Lambda),
 \ee
   has a a quasi-polynomial solution of the form 
\be
\label{16giugno2026-2}   w(x;\Lambda)= \sum_{k=0}^N w_{2k}(\Lambda) x^{-2k}\,x^M\,e^{-\vartheta(x)}.
  \ee
  One can take 
  $$ 
  w_0=0,\quad  w_{2k}(\Lambda)=-\left.\frac{\partial}{\partial\lambda} c_{2k}(\lambda)\right|_{\lambda=\Lambda},
  $$
  where  $c_{2k}(\lambda)$ is  \eqref{6maggio2026-1}. 
   \ele
   
   \bre{\rm Clearly, \eqref{16giugno2026-1} is the non-homogeneous equation associated with \eqref{30aprile2024-2}. 
   If  $w(x;\Lambda)$ is a solution above, so is any $w(x;\Lambda)+\alpha \,y_1(x;\Lambda)$ for any arbitrary constant $\alpha$, and 
   $w_{2k}(\Lambda)\mapsto -\left.\frac{\partial}{\partial\lambda} c_{2k}(\lambda)\right|_{\lambda=\Lambda}+ \alpha\,c_{2k}(\Lambda)$. 
   }\ere

   \begin{proof}
Since the geometric multiplicity is always equal to 1,   $\Lambda$ is eigenvector of $-\mathcal{M}(b,M,N)$ with algebraic multiplicity $\geq 2$ if and only if  there is a principal (or generalized) eigenvector $\boldsymbol{w}$ satisfying 
\be
  \label{23aprile2026-1}
  \Bigl(\mathcal{M}(b,M,N)+\Lambda\Bigr) \boldsymbol{w}= \boldsymbol{c}, 
\quad \quad \boldsymbol{w}=\begin{pmatrix} {w}_{2N}
  \\
  {w}_{2N-2}
  \\
  \vdots
  \\
  {w}_2
  \\
  {w}_0
  \end{pmatrix} \neq 0,
   \ee 
  where $\boldsymbol{c}$ is the eigenvector of $\mathcal{M}$  in \eqref{30aprile2024-5-bis}. 
   Following exactly  the construction in the proof of Proposition \ref{13aprile2023-4}, we see that  \eqref{23aprile2026-1} is equivalent to the fact that the non-homogeneus equation \eqref{16giugno2026-1} has solution \eqref{16giugno2026-2}.
     
To obtain the expression for the coefficients $w_{2k}$, we differentiate w.r.t.  $\lambda$ the equality \eqref{3giugno2026-1} end multiply to the right by the first standard column vector $\boldsymbol{e}_1$. We receive
\be
\label{22aprile2026-5}
  (\mathcal{M}(b,M,N)+\lambda)
  \,\,
  \partial_\lambda \left[\bigl(\mathrm{adj}(\mathcal{M}(b,M,N)+\lambda)\bigr)\,\boldsymbol{e}_1\right]+\mathrm{adj}(\mathcal{M}(b,M,N)+\lambda)\,\boldsymbol{e}_1=
  \chi^\prime_{N+1}(\lambda) \,\boldsymbol{e}_1,
\ee
where $\chi^\prime_{N+1}(\lambda) := \partial_\lambda \chi_{N+1}(\lambda)$. 
    If  the algebraic multiplicity of $\Lambda$ is greater than one, that is $\chi^\prime_{N+1}(\Lambda)=0$, then 
 \eqref{22aprile2026-5} becomes 
  $$ 
 \mathrm{adj}(\mathcal{M}(b,M,N)+\Lambda) \boldsymbol{w}_1(\Lambda)=\boldsymbol{v}_1(\Lambda),
 $$
 where $\boldsymbol{v}_1$ is the eigenvector \eqref{16giugno2026-5} and $$
  \boldsymbol{w}_1(\lambda):=- \frac{\partial  \boldsymbol{v}_1(\lambda)}{\partial \lambda}.
  $$ 
 We conclude that  $\boldsymbol{w}_1(\Lambda)$ is a principal eigenvector. Note that the fact that  $\boldsymbol{v}_1(\Lambda)\neq 0$ and that $\chi^\prime_{N+1}(\Lambda)=0$ implies that $\boldsymbol{w}_1(\Lambda)\neq 0$.  
 Taking the rescaling \eqref{16giugno2026-12}, 
    which gives the coefficients $c_{2k}$  of $Q_1(x,\Lambda)$, 
     we receive the  principal  eigenvector
  \be
 \label{22aprile2026-6}
  \boldsymbol{w}=-\left.\frac{\partial \boldsymbol{c}(\lambda)}{\partial\lambda}\right|_{\lambda=\Lambda}.
  \ee
  \end{proof}

  Consider now the case of Proposition \ref{1febbraio2026-1}, with $\gamma=(2N+M+4)(2N+M+3)$.
  The algebraic multiplicity of an eigenvalue $-\Lambda$ of $ i\mathcal{M}( i b, -M-3 ,N)$ is greater than one  if and only if  $b$ is a  root of  
\be
\label{14giugno2025-5-BIS}
{\rm res}_\lambda\Bigl(\det(\mathcal{M}( i b, -M-3 ,N)+\lambda),\,\frac{\partial}{\partial \lambda}\det(\mathcal{M}(i b,-M-3,N)+\lambda)\Bigr)=0
\ee

\bre{\rm

Since the sign in $\pm i\mathcal{M}(\pm i b, -M-3 ,N)$ is irrelevant, throughout the rest of the paper we will simply use  $i\mathcal{M}(i b,-M-3,N)$, in order to simplify the notation.

}
\ere
   The following lemma is completely  analogous to Lemma \ref{3giugno2026-2}, so we omit its proof.

  \ble
  \label{3giugno2026-7}
  Let $M\in \mathbb{C}$, $N\in\mathbb{N}$ and $\gamma=(2N+M+4)(2N+M+3)$ as in Proposition \ref{1febbraio2026-1}, and let  $y_2(x;\Lambda)$ be  the quasi-polynomial solution of \eqref{30aprile2024-2} corresponding to an eigenvalue $-\Lambda$ of $ i\mathcal{M}(b,-M-3,N)$. 
  Then, $\Lambda$  has algebraic multiplicity greater than one  if and only if the non-homogeneus equation
   $$
   \frac{d^2 w}{dx^2}+\left[\Lambda-\left(x^6+2bx^4+(b^2-2M-3)x^2+\frac{\gamma}{x^2}\right)\right]w= y_2(x;\Lambda),
    $$
   has a quasi-polynomial solution  
   $$    \widetilde{w}(x;\Lambda)= \sum_{k=0}^N f_{2k}(\Lambda) x^{-2k}\,x^{-M-3}\,e^{\vartheta(x)}.
   $$
 One can take
   $$ 
   f_0=0,\quad 
f_{2k}(\Lambda)=-\left.\frac{\partial }{\partial \lambda}d_{2k}(\lambda)\right|_{\lambda=\Lambda},
$$
where   $d_{2k}(\lambda)$ is \eqref{6maggio2026-3-bis}.
\ele

%%%%%%%%%%%%%%%%%%%%%
\section{Painlev\'e IV}

 Before proving Theorem \ref{17giugno2025-1}, we need to recall some basic facts concerning the poles of solutions of the fourth Painlev\'e equation (hereafter denoted by PIV), the associated quadratic anharmonic oscillator, and rational solutions.  Equation PIV 
$$
\frac{d^2 u}{dt^2}= \frac{1}{2u}\left(\frac{du}{dt}\right)^2+\frac{3}{2}\,u+4tu^2+2(t^2+1-2\theta_\infty)u-\frac{8\theta_0^2}{u}
$$
is the isomonodromy deformation condition of a linear $2\times 2$ system of differential equations with an independent variable, say $s$, and  an isomonodromic  parameter $t$ (see the seminal paper \cite{JM-2} of Jimbo and Miwa). The $2\times 2$ isomonodromic system can be reduced  to an equivalent  second order scalar differential equation in normal form \cite{MR}
$$\frac{d^2 \psi}{d s^2} = W(s,t) \psi.
$$

 It is also well known that the poles $t=a$ of  a PIV transcendent can have residue $\pm 1$ only, with Laurent expansions depending on two parameters $(a,C_2)$:
$$
u(t)=\frac{1}{t-a}-a+\frac{a^2+4\theta_\infty-6}{3}(t-a)+C_2(t-a)^2+\dots\quad\quad\hbox{ for residue $+1$};
$$
$$
u(t)=\frac{-1}{t-a}-a-\frac{a^2+4\theta_\infty+2}{3}(t-a)+C_2(t-a)^2+\dots\quad\quad\hbox{ for residue $-1$}.
$$
As well studied in \cite{MR}, the coefficient $W(s,t)$ admits limit at a pole $t=a$, and the  limiting equation is a singular\footnote{i.e., with centrifugal term $s^{-2}$.} quadratic anharmonic oscillator of the form   
\be
\label{14giugno2025-3}
\frac{d^2 \psi}{d s^2}=\underbrace{\left( s^2+2As+A^2+B+\frac{\Lambda_1}{s}+\frac{\Gamma}{s^2}\right)}_{ \lim_{t\to a}W(s,t)}\psi,
\ee
 with coefficients
\be
\label{9giugno2026-1}
 \begin{aligned}
&\quad A=a,\quad \Gamma= \theta_0^2-\frac{1}{4},
 \\
 &
 \left\{ 
 \begin{aligned}
 & B=-2\theta_\infty,\quad \Lambda_1=-C_2+a\left(\frac{3}{2}-2\theta_\infty\right) , \quad\quad\hbox{ for residue $+1$};
\\
&
B=2-2\theta_\infty,\quad \Lambda_1=-C_2+a\left(\frac{1}{2}-2\theta_\infty\right), \quad\quad\hbox{ for residue $-1$}.
\end{aligned}
\right.
\end{aligned}
\ee

\subsection{   Formal identification with the sextic oscillator}
\label{17giugno2025-2}
Also the sextic oscillator \eqref{30aprile2024-2} becomes a singular  quadratic anharmonic oscillator as above with the change of variables 
\be
\label{8febbraio2026-4}
s=\frac{x^2}{\sqrt{2}},\quad \psi(s)=x^{1/2} y(x).
\ee
In this case, the eigenvalue problem  \eqref{30aprile2024-2} is rewritten as 
\be
\label{16giugno2025-2}
\frac{d^2 \psi}{d s^2}=\left( 
s^2+2As+A^2+B-\frac{\sqrt{2}\,\lambda}{4\,s}+\frac{\Gamma}{s^2}\right)\psi,
\ee
with 
\be
\label{9giugno2026-2}
A=\frac{b}{\sqrt{2}},\quad \Gamma=\frac{1}{4}\left(\gamma-\frac{3}{4}\right), \quad B=-M-\frac{3}{2}.
\ee
So, if $\lambda=\Lambda$ is an eigenvalue corresponding to a quasi-polynomial solution, we have the  formal identifications between \eqref{14giugno2025-3}-\eqref{9giugno2026-1}, and \eqref{16giugno2025-2}-\eqref{9giugno2026-2} given by 
\be
\label{5giugno2026-1}
b=\sqrt{2} a, \quad \gamma=4\theta_0^2-\frac{1}{4},
\ee
and
$$ 
\theta_\infty=\frac{M}{2}+\frac{3}{4},\quad C_2=a\left(\frac{3}{2}-2\theta_\infty\right)+\frac{\sqrt{2}\Lambda}{4},\quad\quad\hbox{ for residue $+1$};
$$
\be
\label{5giugno2026-2}
\theta_\infty=\frac{M}{2}+\frac{7}{4},\quad C_2=a\left(\frac{1}{2}-2\theta_\infty\right)+\frac{\sqrt{2}\Lambda}{4},\quad\quad\hbox{ for residue $-1$}.
\ee

\subsection{  Poles of rational solutions}
\label{15giugno2026-3}

As anticipated in the introduction,  the zeros and  poles of the rational solutions of PIV are roots of either the generalized Hermite polynomials or Okamoto polynomials. The Noumi-Yamada generalized Hermite polynomials  are defined by \cite{BM,MR}
$$
\begin{aligned}
&H_{0,0}(t)=H_{1,0}(t)=H_{0,1}(t)=1,\quad \quad H_{1,1}(t)=2t,
\\
\noalign{\medskip}
&2mH_{m+1,n}(t)H_{m-1,n}(t)=H_{m,n}(x)H^{\prime\prime}_{m,n}(t)-H^\prime_{m,n}(t)^2+2m H_{m,n}(t),
\\
\noalign{\medskip}
& 2nH_{m,n+1}(t)H_{m,n-1}(t)=-H_{m,n}(t)H^{\prime\prime}_{m,n}(t)+H^\prime_{m,n}(tt)^2+2n H_{m,n}(t),
\end{aligned}
$$
The degree of  $H_{mn}(t)$ is $m\cdot n$. The distribution of the zeros of  $H_{m,n}(t)$ was thoroughly studied in \cite{MR, MR1}. See also \cite{BM}. In figure \ref{14-giugno-2025-2} we have drawn the typical distribution of these zeros.

To the above polynomials is associated a class of  rational solution of Painlev\'e IV, which is divided into type 1, 2 and 3, as follows (the notations are from \cite{BM}). 

\begin{itemize}
\item Type 1. The solutions  
$$ 
u_{\rm gH}^{[1]}(t;m,n):=2n\frac{H_{m,n+1}(t)H_{m+1,n-1}(t)}{H_{m,n}(t)H_{m+1,n}(t)},
\quad \theta_0=\pm \frac{n}{2},\quad\theta_\infty=1+m+\frac{n}{2}.
$$
In the denominator a zero  of  $H_{m,n}(t)$   is a pole of residue $-1$, a zero of  $H_{m+1,n}(t)$   is a pole of residue $+1$.

\item Type 2. The solutions
$$ 
u_{\rm gH}^{[2]}(t;m,n):=-2m\frac{H_{m-1,n+1}(t)H_{m+1,n}(t)}{H_{m,n}(t)H_{m,n+1}(t)},\quad 
\theta_0=\pm \frac{m}{2},\quad\theta_\infty=-n-\frac{m}{2}.
$$
In the denominator a zero of  $H_{m,n+1}(t)$   is a pole of residue $-1$, a zero of  $H_{m,n}(t)$   is a pole of residue $+1$.

\item Type 3. The solutions are
$$ 
u_{\rm gH}^{[3]}(t;m,n):=-\frac{H_{m,n}(t)H_{m+1,n+1}(t)}{H_{m,n+1}(t)H_{m+1,n}(t)},
\quad 
\theta_0=\pm \frac{m+n+1}{2},\quad\theta_\infty=\frac{n-m+1}{2}.
$$
In the denominator a zero of  $H_{m+1,n}(t)$   is a pole of residue $-1$, a zero of  $H_{m,n+1}(t)$   is a pole of residue $+1$
\end{itemize}

 It is worth recalling that $H_{mn}$ and $H_{m^\prime n^\prime}$ do not have common roots for $(m,n)\neq (m^\prime,n^\prime)$.  The following statement, from Theorem 2.2., point H.3, of \cite{MR} will be important for us.

\bth
\label{15giugno2025-8}
\cite{MR}
Fix $m,n\in \mathbb{N}_{\geq 1}$. Then $t = a$ is a zero of the generalised Hermite polynomial $H_{mn}(t)$ if and only if 
there exists $C_2\in\mathbb{C}$  (unique), such that the anharmonic oscillator \eqref{14giugno2025-3}  corresponding to residue $-1$, and with 
$$\theta_0 = \frac{m+n}{2} , \quad  \theta_\infty = \frac{n -m + 2}{2},$$ 
has two linearly independent solutions 
\be
\label{15giugno2025-9} 
\psi_0(s)=P(s)~s^{-(m+n-1)/2} ~e^{-g(s,a)},
\ee
\be
\label{15giugno2025-10}
\psi_1(s)=Q(s)~ s^{-(m+n-1)/2}~e^{g(s,a)},
\ee
where 
$$
g(s,\lambda):=\frac{1}{2}s^2+a s.
$$ 
and $P(s)$ and $Q(s)$ are polynomials of degree $n-1$ and $m-1$ respectively without repeated roots, with $P(0)\neq 0$ and $Q(0)\neq 0$.
\eth

 %%%%%%%%%%%%%%%%%%%%

 \section{Proof of Theorem \ref{17giugno2025-1}}
 
 The above preparation allows us to prove Theorem  \ref{17giugno2025-1}, stated in the Introduction. 
 
 \subsection{A preliminary remark}

Consider the three-diagonal  $(N+1)\times(N+1)=(m+n)\times(m+n)$ matrix $\mathcal{M}(b,M,N) $,  defined in  \eqref{28maggio2024-1}. Its  $k^\mathrm{th}$ rows has three non-zero entries corresponding to columns $k-1$, $k$ and $k+1$ respectively. With parameters  \eqref{16marzo2026-1}, this row is 
\be
\label{16giugno2025-9}
\hbox{row $k$ of $\mathcal{M}$ }
\longrightarrow 
\quad 
\Bigl[
     0, \dots,0, 
      ~\underset{\hbox{column } k-1}{4(m+n-k+1)},    ~\underset{\hbox{column }k}{\sqrt{2}\,a\, (2n+2-4k)},~
        \underset{\hbox{column } k+1}{4k(k-n)},~0,~\dots,0\Bigr],
\ee
where there is no column $k-1$  for $k=1$ and no  column $k+1$ for $k=m+n$. Hence, it satisfies the crucial property $$ 
\mathcal{M}_{n,n+1}=0. 
$$
Let $\mathcal{M}_1$ and $\mathcal{M}_2$ be   the upper-left $n\times n$ block and the lower-right $m \times m$ block  respectively. For example, in case $n=3$ and $m=2$, we have 
$$
\mathcal{M}(\sqrt{2}a)=\left(\begin{array}{ccc|cc}
4 \sqrt{2}\, a  & -8 & 0 & 0 & 0 
\\
 16 & 0 & -8 & 0 & 0 
\\
 0 & 12 & -4 \sqrt{2}\, a  & \boldsymbol{0} & 0 
\\
\hline
 0 & 0 & 8 & -8 \sqrt{2}\, a  & 16 
\\
 0 & 0 & 0 & 4 & -12 \sqrt{2}\, a  
\end{array}
\right),\quad \hbox{ with } \mathcal{M}_{34}=0,\quad\quad b=\sqrt{2}a,
$$
and 
$$
\mathcal{M}_1=\left(\begin{array}{ccc}
4 \sqrt{2}\, a  & -8 & 0 
\\
 16 & 0 & -8 
\\
 0 & 12 & -4 \sqrt{2}\, a  
\end{array}\right),\quad \mathcal{M}_2=\left(\begin{array}{cc}
-8 \sqrt{2}\, a  & 16 
\\
 4 & -12 \sqrt{2}\, a  
\end{array}\right).
$$
 This structure implies that the characteristic polynomial \eqref{21febbraio2026-1} 
is factorized as
\be
\label{5maggio2026-5}
p(\lambda;\sqrt{2}a)=p_1(\lambda;\sqrt{2}a) \cdot p_2(\lambda;\sqrt{2}a),
\ee
where
$$
p_1(\lambda;\sqrt{2}a):=\det(\mathcal{M}_1(\sqrt{2} a)+\lambda I_n),\quad p_2(\lambda;\sqrt{2}a):=\det(\mathcal{M}_2(\sqrt{2}a)+\lambda I_m).
$$
Thus, the set of eigenvalues of $\mathcal{M}$ is the union of those of $\mathcal{M}_1$ and $\mathcal{M}_2$.   Therefore,  there is a natural factorization of \eqref{14giugno2025-5} with three factors. First, if $\mathcal{M}_j$, for  $j=1 $ or $2$, has an eigenvalue of algebraic multiplicity $\geq 2$,  the same applies to  $\mathcal{M}$ . Hence, for some integers $n_1,\,n_2\geq1$, the resultant must have a factor 
$$r_1(a)^{n_1}\,r_2(a)^{n_2}.
$$
 Moreover, if $p_1$ and $p_2$ have a common root, then this at least has algebraic multiplicity two for $p$.  Therefore, for some integer $n_3\geq1$ , the resultant must have a factor 
$$
{\rm res}_\lambda\Bigl( \det(\mathcal{M}_1(\sqrt{2}a)+\lambda I_n), \det(\mathcal{M}_2(\sqrt{2}a)+\lambda I_m)\Bigr)^{n_3}.
$$
We will prove that the above resultant is proportional to $H_{m,n}$, and that   $n_1=n_2=1$ and $n_3=2$.

\subsection{Proof of Theorem \ref{17giugno2025-1}}

{\bf Step 1: factorization of the characteristic polynomial of $\mathcal{M}$.} We have already observed that, with parameters  \eqref{16marzo2026-1}, 
the crucial property is  $$ 
\mathcal{M}_{n,n+1}=0,
$$
so that the factorization \eqref{5maggio2026-5} holds. 
\vskip 0.2 cm 
\noindent
{\bf Step 2: factorization of the resultant.} The factorization $p=p_1p_2$ implies that (see Lemma \ref{1febbraio2026-5} in Appendix 1)
$$ 
{\rm res}_\lambda \left(
p(\lambda),\frac{\partial p(\lambda)}{\partial \lambda}\right)
=(-1)^{mn}
{\rm res}_\lambda \left( p_1,\frac{\partial p_1}{\partial \lambda}\right)
\cdot
{\rm res}_\lambda \left( p_2,\frac{\partial p_2}{\partial \lambda}\right)
\cdot 
\left({\rm res}_\lambda \left( p_1,p_2\right)\right)^2.
$$
This proves that $n_1=n_2=1$ and $n_3=2$. 
\vskip 0.2 cm 
\noindent
{\bf Step 3: identification of ${\rm res}_\lambda \left( p_1,p_2\right)$ with $H_{m,n}$.} 
We  show that
$$
H_{m,n}(a)=c_{mn}\cdot {\rm res}_\lambda \left( p_1(\lambda;\sqrt{2}a),p_2(\lambda;\sqrt{2}a)\right),
$$
where $c_{mn}\neq 0$ is a constant.  
In order to do this, we make use of the result of \cite{MR}, expressed in Theorem \ref{15giugno2025-8}. The two  functions  \eqref{15giugno2025-9} and \eqref{15giugno2025-10} are both solutions of the oscillator \eqref{14giugno2025-3}. 
In analogy with \eqref{16giugno2025-2}, that represents the eigenvalue problem \eqref{30aprile2024-2}, 
it will be convenient to introduce the "eigenvalue" $\Lambda_2$ (analogous of $\Lambda$) by
$$
\Lambda_1=:-\frac{\sqrt{2}\, {\Lambda_2}}{4 }, \quad\hbox{ that is } C_2=a\left(\frac{1}{2}-2\theta_\infty\right)+\frac{\sqrt{2}\Lambda_2}{4}
$$
and rewrite  \eqref{14giugno2025-3} as 
\be
\label{16giugno2025-3}
\frac{d^2 \psi}{d s^2}=\left(s^{2}+2 a s +(a^{2}+2-2 \theta)  -\frac{\sqrt{2}\, \Lambda_2}{4 s}+\frac{\theta_0^{2}-\frac{1}{4}}{s^{2}}\right)\psi,\quad \theta_0 = \frac{m+n}{2} , \quad  \theta_\infty = \frac{n -m + 2}{2}.
\ee
First, consider the solution \eqref{15giugno2025-9}:
$$\psi_0(s)=P(s)~s^{-(m+n-1)/2} ~e^{-g(s,a)},
\quad\quad P(s)=\sum_{k=0}^{n-1} P_k s^k. 
$$
The direct substitution of  \eqref{15giugno2025-9} into the anharmonic oscillator \eqref{16giugno2025-3}  gives the following  recurrence relations  (the equation for $k=0$ is identically satisfied):
\be
\label{9giugno2026-3}
\begin{aligned}
& \left(1-m -n \right)P_{1} +a  \left(m +n -1\right)P_{0} = 
-\frac{\sqrt{2}\, \Lambda_2}{4} P_{0},\quad \hbox{for $k=1$},
\\
&
k  \left(k -m -n \right)P_{k}-a  \left(2 k -1-m -n \right)P_{k -1}-2  \left(k -1-n \right)P_{k-2}
 = -\frac{\sqrt{2}\, \Lambda_2  }{4}P_{k -1},\quad \hbox{ for $k\geq 2$.}
 \end{aligned}
 \ee
 The coefficient  $-2(k - 1 - n)$  of $P_{k-2}$ vanishes for $k=n+1$. This implies that,  for a chosen initial value $P_0$, it is possible to solve the recurrence uniquely with $P_j=0$  for all $j\geq n$.\footnote{ The relations for $k=n-1,n, n+1$ are
 $$
 \begin{aligned}
 &\left(n -1\right) \left(-1-m \right)P_{n -1} -a  \left(n -3-m \right)P_{n-2}+4P_{n-3}
 = -\frac{\sqrt{2}\, \Lambda_2\,  P_{n-2}}{4}
\\
&
 -mn P_{n}  -a \left(n -1-m \right) P_{n -1}+2 P_{n-2} = 
-\frac{\sqrt{2}\, \Lambda_2  P_{n -1}}{4};
\\
&
\left(1-m \right)\left(n +1\right) P_{n +1} -a \left(n +1-m \right)P_{n} 
 = -\frac{\sqrt{2}\, \Lambda_2  P_{n}}{4}.
\end{aligned}$$
It is possible to take $P_n=P_{n+1}=0$. The first line determines $P_{n-1}$ from previous steps. The second line is satisfied being equivalent to  $\det(\mathcal{N}^{(0)}+\Lambda_2)=0$ and the third line is $0=0$. All the other successive relations will be identically satisfied by $P_j=0$ for all $j\geq n$.
}
 The recurrence relations are equivalent to 
 \be
 \label{16giugno2025-5} 
 \mathcal{N}_0(a)\begin{pmatrix}
 P_0
 \\
 P_1
 \\
 \vdots
 \\
 P_{n-1}
 \end{pmatrix}= -\Lambda_2
 \begin{pmatrix}
 P_0
 \\
 P_1
 \\
 \vdots
 \\
 P_{n-1}
 \end{pmatrix}
 ,\quad \hbox{ and } \det(\mathcal{N}_0(a)+\Lambda_2)=0,
 \ee
 which is an eigenvalue problem for a certain  tridiagonal $n\times n$ matrix  $\mathcal{N}_0(a)$, that we will explicitly  write below (see \eqref{16giugno2025-10}). 
 
 Successively, we consider the solution \eqref{15giugno2025-10}:
 $$
 \psi_1(s)=Q(s)~ s^{-(m+n-1)/2}~e^{g(s,a)},
\quad\quad 
 Q(s)=\sum_{k=0}^{m-1} Q_k s^k.
 $$
 Repeating a construction analogous to the above, we find  the recurence relations
\be
\label{9giugno2026-4}
 \begin{aligned}
&  \left(1-m -n \right)Q_{1}+a \left(1-m -n \right)Q_{0} =-\frac{ \sqrt{2}\, \Lambda_2  }{4}\,Q_{0},\quad k=1,
\\
&
  k \left(k -m -n \right)\,Q_{k}+a  \left(2 k -1-m -n \right)\,Q_{k -1}+2  \left(k -1-m \right)\,Q_{-2+k}=-\frac{ \sqrt{2}\, \Lambda_2}{4}\,Q_{k -1},\quad k\geq 2.
 \end{aligned}
 \ee
Since $k -1-m =0$ for $k=m+1$, the recurrence has solution $(Q_0,Q_1,...,Q_{m-1},\,0, \,0,\,....)$, with all $Q_j=0$ for $j\geq m$.  The above is equivalent to an eigenvalue problem 
\be
 \label{16giugno2025-6} 
 \mathcal{N}_1(a)\begin{pmatrix}
 Q_0
 \\
 Q_1
 \\
 \vdots
 \\
 Q_{m-1}
 \end{pmatrix}= -\Lambda_2
 \begin{pmatrix}
 Q_0
 \\
 Q_1
 \\
 \vdots
 \\
 Q_{m-1}
 \end{pmatrix}
 ,\quad \hbox{ and } \det(\mathcal{N}_1(a)+\Lambda_2)=0,
 \ee
for a certain  tridiagonal $m\times m$ matrix  $\mathcal{N}_1(a)$.  Explicit forms will be   writen below (see \eqref{16giugno2025-12}) . 
  
  \vskip 0.2 cm
  Theorem \ref{15giugno2025-8} is now rephrased as follows: $t=a$ is a zero of $H_{mn}(t)$ if and only if there exists $C_2$ -- that is an eigenvalue $\Lambda_2$ -- such that both eigenvalue problems \eqref{16giugno2025-5} and  \eqref{16giugno2025-6}, with the given $a$, have solution with {\it the same} $\Lambda_2$.  Equivalently,  $t=a$ is a zero of $H_{mn}(t)$ if and only if $ \det(\mathcal{N}_0(a)+\lambda)$ and $\det(\mathcal{N}_1(a)+\lambda)$ have a common root, that is  
  $$ 
  {\rm res}_\lambda\Bigl(\det(\mathcal{N}_0(a)+\lambda),\,\det(\mathcal{N}_1(a)+\lambda)\Bigr)=0.
  $$
  The recurrence relations are linear of $a$, hence the l.h.s. above is a polynomial in $a$ of degree $m\cdot n$. On the other hand, also the degree of  $H_{mn}(a)$ is $m\cdot n$. Therefore, 
   by Theorem  \ref{15giugno2025-8},  ${\rm res}_\lambda\Bigl(\det(\mathcal{N}_0(a)+\lambda),\,\det(\mathcal{N}_1(a)+\lambda)\Bigr)$ and $H_{mn}(a)$ must have the same roots, so there is a constant $c_{mn}\neq 0$ such that  
  $$
  H_{mn}(a)=c_{mn} \,{\rm res}_\lambda\Bigl(\det(\mathcal{N}_0(a)+\lambda),\,\det(\mathcal{N}_1(a)+\lambda)\Bigr).
  $$
  
  \vskip 0.2 cm 
  Instead of the matrices $\mathcal{N}_0$ and $\mathcal{N}_1$, it will be convenient to consider the matrices 
  $$
   \mathcal{B}_0:=-(\mathcal{N}_0+2a\sqrt{2}\,m\,I_n),\quad   \mathcal{B}_1:=-(\mathcal{N}_1+2a\sqrt{2}\,m\,I_m).
   $$
   The eigenvalue problems \eqref{16giugno2025-5} and \eqref{16giugno2025-6}  have the same eigenvalue $2a\sqrt{2}\,m-\Lambda_2$. The resultant is unchanged, so that
  $$
  H_{mn}(a)=c_{mn}\, {\rm res}_\lambda\Bigl(\det(\mathcal{B}_0(a)+\lambda),\,\det( \mathcal{B}_1(a)+\lambda)\Bigr).
  $$
   In place of $ \mathcal{B}_1$ we write the reverted matrix $\mathcal{A}_1$ defined by reverting the order of the entries, that is 
$$ 
(\mathcal{A}_1)_{j\ell}:=( \mathcal{B}_1)_{m-j+1,m-\ell+1}.
$$
Our eigenvalue problem 
$$  \mathcal{B}_1\begin{pmatrix}
 Q_0
 \\
 Q_1
 \\
 \vdots
 \\
 Q_{m-1}
 \end{pmatrix}=(2a\sqrt{2}\,m-\Lambda_2)
 \begin{pmatrix}
 Q_0
 \\
 Q_1
 \\
 \vdots
 \\
 Q_{m-1}
 \end{pmatrix}
 $$
 becomes
$$ \mathcal{A}_1\begin{pmatrix}
 Q_{m-1}
 \\
 Q_{m-2}
 \\
 \vdots
 \\
 Q_0
 \end{pmatrix}=(2a\sqrt{2}\,m-\Lambda_2)
 \begin{pmatrix}
 Q_{m-1}
 \\
 Q_{m-2}
 \\
 \vdots
 \\
 Q_0
 \end{pmatrix}.
 $$
The resultants do not change, so that   
$$
  H_{mn}(a)=c_{mn} \,{\rm res}_\lambda\Bigl(\det(\mathcal{B}_0(a)+\lambda),\,\det(\mathcal{A}_1(a)+\lambda)\Bigr).
  $$
From the recurrence relations \eqref{9giugno2026-3} and \eqref{9giugno2026-4}, we can write the explicit form of the matrices:
\be
\label{16giugno2025-10}
\mathcal{B}_0:\quad 
\hbox{row $k$ }
\longrightarrow 
\quad 
\Bigl[
     0, \dots,0, 
      ~\underset{\hbox{column } k-1}{4\sqrt{2}\,(n+1-k)},    ~\underset{\hbox{column }k}{\sqrt{2}a \,(2n+2-4k)},~
        \underset{\hbox{column } k+1}{-2\sqrt{2}\,k(m+n-k)},~0,~\dots,0\Bigr],
\ee
where $k=1,...,n$ (there is no column $k-1$  for $k=1$ and no  column $k+1$ for $k=n$).

 \be
 \label{16giugno2025-12}
\mathcal{A}_1:\quad 
\hbox{row $k$ }
\longrightarrow 
\quad 
\Bigl[
     0, \dots,0, 
      ~\underset{\hbox{column } k-1}{2\sqrt{2}\,(m-k+1)(1-n-k)},    ~\underset{\hbox{column }k}{\sqrt{2}a\, (2-2n-4k)},~
        \underset{\hbox{column } k+1}{-4\sqrt{2}\,k},~0,~\dots,0\Bigr],
\ee
where $k=1,...,m$ (there is no column $k-1$  for $k=1$ and no  column $k+1$ for $k=m$).
\vskip 0.2 cm 
The proof of the theorem is completed by observing that  
$$ 
{\rm res}_\lambda \left( p_1(\lambda;a),p_2(\lambda;a)\right)={\rm res}_\lambda\Bigl(\det(\mathcal{B}_0(a)+\lambda),\,\det(\mathcal{A}_1(a)+\lambda)\Bigr)
$$
 This follows from the fact that  $\mathcal{B}_0$ is similar to $\mathcal{M}_1$, and   $\mathcal{A}_1$ is similar to $\mathcal{M}_2$.  We prove the similarity.

\vskip 0.2 cm
i)    $\mathcal{B}_0$ and $\mathcal{M}_1$ are similar:  the similarity is realized by a diagonal matrix
$$ 
\mathrm{diag}(\alpha_1,\cdots,\alpha_n)^{-1}\, \mathcal{B}_0\,\mathrm{diag}(\alpha_1,\cdots,\alpha_n)= \mathcal{M}_1.
$$
Indeed, the 
$k^\mathrm{th}$ row of $\mathrm{diag}(\alpha_1,\cdots,\alpha_n)^{-1}\, \mathcal{B}_0\,\mathrm{diag}(\alpha_1,\cdots,\alpha_n)$ is 
$$\Bigl[
     0, \dots,0, 
      ~\underset{\hbox{column } k-1}{4\sqrt{2}(n+1-k)}\,\frac{\alpha_{k-1}}{\alpha_k},    ~\underset{\hbox{column }k}{\sqrt{2}a\, (2n+2-4k)},~
        \underset{\hbox{column } k+1}{-2\sqrt{2}\,k(m+n-k)}\,\frac{\alpha_{k+1}}{\alpha_k},~0,~\dots,0\Bigr],\quad k=1,...,n,
$$
so that comparison with $\mathcal{M}_1$, whose rows are the first $n$ rows of \eqref{16giugno2025-9} (up to the $n^\mathrm{th}$ column), shows that the similarity is  realized by   $\alpha_1,...,\alpha_n$ satisfying
$$ 
\sqrt{2}\,(n-k)\,\frac{\alpha_k}{\alpha_{k+1}}=(m+n-k),\quad k=1,...,n-1.
$$

ii)  $\mathcal{A}_1$ and $\mathcal{M}_2$ are similar, by
$$ 
\mathrm{diag}(\beta_1,\cdots,\beta_m)^{-1}\, \mathcal{A}_1\,\mathrm{diag}(\beta_1,\cdots,\beta_m)= \mathcal{M}_2.
$$
To see this, observe that 
the $k^\mathrm{th}$ row of $\mathrm{diag}(\beta_1,\cdots,\beta_m)^{-1}\, \mathcal{A}_1\,\mathrm{diag}(\beta_1,\cdots,\beta_m)$ is 
$$ 
\Bigl[
     0, \dots,0, 
      ~\underset{\hbox{column } k-1}{2\sqrt{2}\,(m-k+1)(1-n-k)\,\frac{\beta_{k-1}}{\beta_k}},    ~\underset{\hbox{column }k}{\sqrt{2}a (2-2n-4k)},~
        \underset{\hbox{column } k+1}{-4\sqrt{2}\,k\,\frac{\beta_{k+1}}{\beta_k} },~0,~\dots,0\Bigr],\quad k=1,...,m,
        $$
The rows of $\mathcal{M}_2$ are obtained form the last $m$ rows of  \eqref{16giugno2025-9}. The $k^\mathrm{th}$ row of $\mathcal{M}_2$ is 
 $$ 
\Bigl[
     0, \dots,0, 
      ~\underset{\hbox{column } k-1}{4(m-k+1)},    ~\underset{\hbox{column }k}{\sqrt{2}a\, (2-2n-4k)},~
        \underset{\hbox{column } k+1}{-4\,k\,(n+k)},~0,~\dots,0\Bigr],\quad k=1,...,m.
$$
Comparison shows that the similarity can be realized by $\beta_1,...,\beta_m$ solving the system
$$
\frac{\beta_k}{\beta_{k+1}}=-\frac{\sqrt{2}}{n+k}.
$$

 \section{On the algebraic spectrum with two simultaneous eigenfunctions}

We investigate the conditions on $b$ and on the eigenvalues $\Lambda$ such that the sextic oscillator has simultaneously two quasi-polynomial solutions with negative and positive exponential factors respectively, which means two opposite behaviours at $\infty$.  We show that also in this case the parameter  $b$ in equation \eqref{30aprile2024-2} must be exactly equal to a  root  $a$ of a generalized Hermite polynomial, upon rescaling $a=b/\sqrt{2}$. 

\subsection{Simultaneous eigenfinctions and Hermite polynomials}

 \ble
\label{6febbraio2026-1}
Equation \eqref{30aprile2024-2} has  simultaneously  two quasi-polynomial solutions $y_1(x,\Lambda)=Q_1(x,\Lambda) e^{-\vartheta(x)}$ and $y_2(x,\Lambda)=Q_2(x,\Lambda) e^{\vartheta(x)}$ of Proposition \ref{13aprile2023-4} and Proposition \ref{1febbraio2026-1} with finite sums having respectively $N_1+1$ and $N_2+1$ terms (with $N_1,N_2\in\mathbb{N}$) if and only if  the following conditions hold.

\begin{itemize}
\item[a)] 
For some $N_1,\,N_2\in\mathbb{N}$ 
\be
\label{7maggio2026-1}
M=N_1-N_2-\frac{3}{2}
,\quad\quad 
 \gamma= \Bigl(N_1 + N_2 + \frac{3}{2}\Bigr) \Bigl(N_1 + N_2 + \frac{5}{2}\Bigr).
\ee
 \item[b)] 
 $b$ is a root of the resultant
 $$ 
 {\rm res}_\lambda\Bigl( \det\bigl(\mathcal{M}( b,N_1-N_2-3/2,N_1)+\lambda\bigr),\,\det\bigl( i {\mathcal{M}}( i b,N_2-N_1-3/2,N_2)+\lambda\bigr)\Bigr)=0,
 $$
 and   $-\Lambda$ is common eigenvalue of the matrices 
 \be
 \label{4giugno2026-1}
 \mathcal{M}\left(b,N_1-N_2-
 {3}/{2},N_1\right) \quad\hbox{ and } \quad   i\mathcal{M}\left( i b,N_2-N_1-
 {3}/{2},N_2\right).
\ee
 \end{itemize}
 Signs $\pm $ in the above formulae give the same result. 
 \ele
 
 \bre
 \label{5giugno2026-8}
  {\rm  By the identities  
\eqref{5giugno2026-5} and \eqref{5giugno2026-7}, the matrices \eqref{4giugno2026-1} can be replaced by the matrices \be
\label{4giugno2026-3}
\mathcal{C}_{N_1+1}\left( b,N_1-N_2-{3}/{2}\right) \quad\hbox{ and } \quad \mathcal{D}_{N_2+1}\left( b,N_1-N_2-{3}/{2}\right).
\ee
}
\ere

 \begin{proof}
 From Propositions \ref{13aprile2023-4} and  \ref{1febbraio2026-1} it follows that the necessary and sufficient condition to have simultaneously  $y_1(x,\Lambda)$ and $y_2(x,\Lambda)$ is the following conditions:  

\vskip 0.2 cm 
\noindent
1) $\gamma$ must satisfy
$$ 
\gamma=  (2N_1-M)(N_1+1-M)=(2N_2+3+M)(2N_2+4+M)$$
which is possible if and only if $M=N_1-N_2-3/2$.  This proves a). 
\vskip 0.2 cm 
\noindent
2) For $M=N_1-N_2-3/2$ as above, the eigenvalue problems associated with $y_1$ and $y_2$ respectively 
$$
(\mathcal{M}(b,M,N_1)+\lambda)\boldsymbol{c}
 = 0
,\quad 
\quad
 ( i {\mathcal{M}}( i b,-M-3,N_2)+{\lambda})
 \widetilde{\boldsymbol{c}}
= 0
$$
must admit a common solution $\lambda=\Lambda$. This proves b). \end{proof}

 \bpr
 \label{8febbraio2026-5}
The sextic oscillator \eqref{30aprile2024-2} has   simultaneously quasi-polynomial solutions $y_1(x,\Lambda)=Q_1(x,\Lambda) e^{-\vartheta(x)}$ and $y_2(x,\Lambda)=Q_2(x,\Lambda) e^{\vartheta(x)}$ of Proposition \ref{13aprile2023-4} and Proposition \ref{1febbraio2026-1} respectively, with finite sums having respectively $N_1+1$ and $N_2+1$ terms, if and only if the following conditions hold
\begin{itemize}
 \item[A)] For some $N_1$, $N_2\in\mathbb{N}$, conditions \eqref{7maggio2026-1} hold.

 \item[B)]  Setting  $a=b/\sqrt{2}$, then $a$ is a  root of 
 $$H_{N_2+1, N_1+1}(a)=0,
 $$
 and   $-\Lambda$ is common eigenvalue of the matrices \eqref{4giugno2026-1} (equivalently, of the matrices \eqref{4giugno2026-3}). \end{itemize}

\epr
 
 \begin{proof} Condition A) is as condition a) of Lemma \ref{6febbraio2026-1}.

\vskip 0.15 cm 
In order to prove B), we preliminarily show that the formal identification \eqref{5giugno2026-1}-\eqref{5giugno2026-2} between the sextic oscillator and the anharmonic oscillator  \eqref{14giugno2025-3} associated with PIV  becomes  an actual identification between a sextic oscillator satisfying condition a) of Lemma \ref{6febbraio2026-1} and an anharmonic oscillator  associated with PIV corresponding to a pole of  residue $-1$ and parameters
 \be
 \label{5giugno2026-3}
 \theta_0 = \frac{m+n}{2} , \quad  \theta_\infty = \frac{n -m + 2}{2}.
 \ee
   First, consider  \eqref{14giugno2025-3} for PIV, corresponding to a pole of residue $-1$ and parameters \eqref{5giugno2026-3}. 
 The identification 
$$
 \theta_\infty=\frac{M}{2}+\frac{7}{4}
$$
 implies
 $$ 
 M= n-m-\frac{3}{2}.
 $$
 Then, the identification 
 $$
 \frac{1}{4}\left(\gamma-\frac{3}{4}\right)=\frac{(m+n)^2-1}{4}
 $$
yields
 $$ 
 \gamma=\frac{(2m + 2n + 1)(2m + 2n - 1)}{4}=\left\{
 \begin{aligned}
 & \underset{n=m+M+\frac{3}{2}}=\Bigl( 2(m-1)+M+3 \Bigr)\Bigl(2(m-1)+M+4\Bigr) 
 \\
 \noalign{\medskip}
 &\underset{m=n-M-\frac{3}{2}}=\Bigl( 2(n-1)-M \Bigr)\Bigl(2(n-1)-M+1\Bigr) 
  \end{aligned}
 \right.
 $$
 We conclude that if we define 
  \be
  \label{8febbraio2026-3}
  N_1=n-1,\quad\quad N_2=m-1,
\ee
then 
$$M=N_1-N_3-\frac{3}{2},\quad\quad  \gamma=(2N_2+M+3)(2N_2+M+4)=(2N_1-M)(2N_1-M+1),
$$ so that  condition a) of Lemma \ref{6febbraio2026-1} is satisfied.  
Conversely, for the sextic oscillator satisfying condition a) of Lemma \ref{6febbraio2026-1}, define $n$ and $m$ using   \eqref{8febbraio2026-3}, so that $M=n-m-3/2$. The identification  
$$
\gamma=4\theta_0^2-\frac{1}{4},\quad  \theta_\infty=\frac{M}{2}+\frac{7}{4}
$$
yields 
$$
\theta_0 = \frac{m+n}{2} , \quad  \theta_\infty = \frac{n -m + 2}{2},
$$
as we wanted to show. 
  
\vskip 0.2 cm 
We are ready to prove B), observing that Theorem  \ref{15giugno2025-8} holds here,  because we are dealing with  residue $-1$ and parameters \eqref{5giugno2026-3}. Now, the solutions in Theorem \ref{15giugno2025-8}  given by 
$$ 
\psi_0(s)=P(s)~s^{-(m+n-1)/2} ~e^{-g(s,a)},
\quad 
\psi_1(s)=Q(s)~ s^{-(m+n-1)/2}~e^{g(s,a)},
$$
  exactly correspond to two solutions $y_1(x)=Q_1(x)e^{-\vartheta(x)}$ and $y_2(x)=Q_2(x)e^{\vartheta(x)}$, because (using that $P$ is a polynomial of degree $n-1$ and $Q$ of degree $m-1$) 
$$ 
\psi_0(s)=  \sum_{k=0}^{n-1} v_k s^{-k} \,s^{(n-m-1)/2}\, e^{-g(s,a)},\quad v_0\neq 0
$$
$$
\psi_1(s)= \sum_{k=0}^{m-1} w_k s^{-k} \,s^{(m-n-1)/2}\, e^{g(s,a)},\quad w_0\neq 0, 
$$
for some coefficients $v_k$ and $w_k$. 
Then by \eqref{8febbraio2026-4} and the identification we are considering,  the above are respectively solutions  of the sextic oscillator of the form 
$$
\begin{aligned}  
& y_1(x)=Q_1(x) e^{-\vartheta(x,b)},\quad Q_1(x)= \sum_{k=0}^{N_1} c_{2k}x^{-2k} ~ x^M, \quad c_0\neq 0
\\
&y_2(x)=Q_2(x) e^{\vartheta(x,b)},\quad Q_2(x)=\sum_{k=0}^{N_2} d_{2k} x^{-2k}\cdot x^{-M-3}, \quad {d}_0\neq 0,
\end{aligned}
$$
with $\vartheta(x,b)=x^4/4+bx^2/2$ and $b=\sqrt{2}\,a$. 
By  Theorem \ref{15giugno2025-8}, the anharmonic oscillator has the two solutions $\psi_0$ and $\psi_1$, with the same unique $C_2$, if and only if $a=b/\sqrt{2}$ is a root of  
$$H_{mn}(a)=0,\quad\quad m=N_2+1,\quad n=N_1+1.
$$
The coefficient $C_2$ corresponds to the common eigenvalue $-\Lambda$ through
$$
C_2=a\left(\frac{1}{2}-2\theta_\infty\right)+\frac{\sqrt{2}\Lambda}{4}.
$$  
\end{proof}

 \bcr
There is a constant $k_{mn}\neq 0$ such that for $m,n\geq 1$
 $$ 
 \begin{aligned}
 k_{mn} &\,H_{mn}(a)=
 \\
 &=
 {\rm res}_\lambda\left( \det\left(\mathcal{M}\left( \sqrt{2}a,n-m-\frac{3}{2},n-1\right)+\lambda\right),\,\det\left( i {\mathcal{M}}\left( i \sqrt{2}a,m-n-\frac{3}{2},m-1\right)+\lambda\right)\right).
 \end{aligned}
 $$

 \ecr

 \begin{proof}
 By Lemma \ref{6febbraio2026-1} and Proposition \ref{8febbraio2026-5},  $H_{N_2+1, N_1+1}(b/\sqrt{2})$ must be  proportional to the resultant 
 $$ 
 {\rm res}_\lambda\Bigl( \det\bigl(\mathcal{M}( b,N_1-N_2-3/2,N_1)+\lambda\bigr),\,\det\bigl(i {\mathcal{M}}( i b,N_2-N_1-3/2,N_2)+\lambda\bigr)\Bigr).
 $$
Moreover, $N_1=n-1$, $N_2=m-1$. 
 \end{proof}

 \subsection{Further comments  on the two facets of the Sextic-Hermite correspondence}
 At the end of Section \ref{28giugno2026-1} we discussed the relationship between Proposition \ref{8febbraio2026-5} and Theorem \ref{17giugno2025-1}, as two  facets of the correspondence between the Sextic oscillator and generalized Hermite polynomials.

 We now provide a more detailed comparison of the two situations, showing  the natural identification between the   three polynomials $R_1$, $R_2$ and $R_3$ described at the end of  Section \ref{28giugno2026-1}  and   $r_1$, $r_2$ and $H_{mn}$ of  Theorem \ref{17giugno2025-1}.  .
 \vskip 0.2 cm 
 
   If $N_1,\,N_2\in\mathbb{N}$, then 
    \be
 \label{8maggio2026-8}
    M:=N_1+2N_2+\frac{3}{2},\quad N:=N_1+N_2+1
    \ee
    satisfy the condition \eqref{16marzo2026-1} of Theorem \ref{17giugno2025-1}.    Conversely, any half-integer $M$ and $N\in \mathbb{N}$ satisfying \eqref{16marzo2026-1} can be represented as \eqref{8maggio2026-8} with $N_1,\,N_2\in\mathbb{N}$. It suffices to take 
    $$
    N_1 =   2N-M - \frac{1}{2},\quad  N_2 =M -N - \frac{1}{2}. 
    $$
    With this identification, in Theorem \ref{17giugno2025-1} we have 
$$ 
m_{\rm Th \,\ref{17giugno2025-1}}=N_2+1,\quad n_{\rm Th \,\ref{17giugno2025-1}}=N_1+1.
$$
Let $$\mathcal{M}_1=\mathcal{M}_1(b,M,N), \quad\hbox{and}\quad \mathcal{M}_2=\mathcal{M}_2(b,M,N)$$ be the  upper-left $n\times n = (N_1+1)\times (N_1+1)$ block and the lower-right $m\times m = (N_2+1)\times (N_2+1) $ block respectively of the matrix $\mathcal{M}(b,\,M,\,N)$, as defined in Theorem \ref{17giugno2025-1}. 

\vskip 0.2 cm  
We also consider the sextic oscillator admitting two simultaneous quasi-polynomial solutions (conditions of Proposition \ref{8febbraio2026-5}):

\begin{itemize}
\item[--]
the solution  $y_1(x,\lambda)=Q_1(x,\lambda) e^{-\vartheta(x)}$ with $N_1+1$ terms and associated matrix $\mathcal{M}(b,N_1-N_2-3/2,N_1)$ ;
\item[--] 
 the solution  $y_2(x,\lambda)=Q_2(x,\lambda) e^{\vartheta(x)}$ with $N_2+1$ terms  and associated matrix $\pm i\mathcal{M}(\pm ib,\,N_2-N_1-3/2,\, N_2)$.
 \end{itemize}

In a way completely analogous to step 3 in the proof of  Theorem \ref{17giugno2025-1}, we can prove the following similarities. 

i) Take $\mathcal{M}\Bigl(b,N_1-N_2-\frac{3}{2},N_1\Bigr)$  and the shift 
$$
\mathcal{B}:=\mathcal{M}\Bigl(b,N_1-N_2-\frac{3}{2},N_1\Bigr)-2b(N_2+1)\, I_{N_1+1}.
$$ 
Then, by a suitable diagonal similarity, one verifies that
$$  
\mathcal{M}_1(b,M,N)=\hbox{diag}(\alpha_1,\dots,\alpha_{N_1+1})^{-1}\cdot \mathcal{B} \cdot \hbox{diag}(\alpha_1,\dots,\alpha_{N_1+1})
$$

ii) Take $\pm i\mathcal{M}\Bigl(\pm i b,N_2-N_1-\frac{3}{2},N_2\Bigr)$ and the shift (the signs $\pm$ will be unessential for the final result) 
$$
\mathcal{B}^{\pm}_1:=\pm i\mathcal{M}\Bigl(\pm i b,N_2-N_1-\frac{3}{2},N_2\Bigr) -2b(N_2+1)\, I_{N_2+1}.
$$ 
Next,  consider the reverted matrix  with entries $(j,k)$ given by
$$
(\mathcal{A}^{\pm})_{jk}:=(\mathcal{B}^{\pm}_1)_{N_2+2-j,\,N_2+2-k},\quad\quad j,k=1,...,N_2+1.
$$
Then, by a suitable diagonal similarity, one verifies  that 
$$
\mathcal{M}_2(b,M,N)=\hbox{diag}(\beta^{\pm}_1,\dots,\beta^{\pm}_{N_2+1})^{-1}\cdot \mathcal{A}^{\pm}\cdot \hbox{diag}(\beta^{\pm}_1,\dots,\beta^{\pm}_{N_2+1})
$$

It is crucial to notice that the shifts above change in both cases i) and ii) the eigenvalues by the same quantity $2b(N_2+1)$. This implies that 
$$\begin{aligned}
{\rm res}_\lambda\left(\mathcal{M}\Bigl(b,N_1-N_2-\frac{3}{2},N_1\Bigr),\partial_\lambda\mathcal{M}\Bigl(b,N_1-N_2-\frac{3}{2},N_1\Bigr)\right)
\\
\noalign{\medskip}
={\rm res}_\lambda(\mathcal{B},\partial_\lambda\mathcal{B})={\rm res}_\lambda(\mathcal{M}_1(b,M,N),\partial_\lambda\mathcal{M}_1(b,M,N));
\end{aligned}
$$

\vskip 0.3 cm 
$$\begin{aligned}
{\rm res}_\lambda\left(\pm i\mathcal{M}\Bigl(\pm i b,N_2-N_1-\frac{3}{2},N_2\Bigr),\partial_\lambda\left(\pm i\mathcal{M}\Bigl(\pm i b,N_2-N_1-\frac{3}{2},N_2\Bigr)\right)\right)
\\
\noalign{\medskip}
={\rm res}_\lambda(\mathcal{A}^{\pm},\partial_\lambda\mathcal{A}^{\pm})={\rm res}_\lambda(\mathcal{M}_2(b,M,N),\partial_\lambda\mathcal{M}_2(b,M,N));
\end{aligned}
$$

\vskip 0.3 cm 
$$
\begin{aligned}
{\rm res}_\lambda\left(\mathcal{M}\Bigl(b,N_1-N_2-\frac{3}{2},N_1\Bigr),\pm i\mathcal{M}\Bigl(\pm i b,N_2-N_1-\frac{3}{2},N_2\Bigr)\right)
\\
\noalign{\medskip}
={\rm res}_\lambda(\mathcal{B},\mathcal{A}^{\pm})={\rm res}_\lambda(\mathcal{M}_1(b,M,N),\mathcal{M}_2(b,M,N)).
\end{aligned}
$$

Setting 
$
a={b}/\sqrt{2}
$
it follows that the factorization \eqref{15giugno2025-2S-BIS} of Theorem \ref{17giugno2025-1} reads
$$
{\rm res}_\lambda\left(\det\Bigl(\mathcal{M}(b,M,N)+\lambda\Bigr),~ \frac{\partial }{\partial \lambda}\det\Bigl(\mathcal{M}(b,M,N)+\lambda\Bigr)\right)=
$$
$$
 =
 (-1)^{(N_2+2)\cdot(N_1+1)}\,\,c_{(N_2+2),(N_1+1)}^{-1}\, \cdot  r_1(a) \cdot r_2(a) \, \cdot H_{N_2+1,N_1+1}(a)^2,
$$
where 
$$ 
\begin{aligned}
&r_1(a)={\rm res}_\lambda\left(\det\left(\mathcal{M}\Bigl(b,N_1-N_2-\frac{3}{2},N_1\Bigr)+\lambda\right), \frac{\partial }{\partial \lambda} \det\left(\mathcal{M}\Bigl(b,N_1-N_2-\frac{3}{2},N_1\Bigr)+\lambda\right)\right),
\\
\noalign{\medskip}
&
r_2(a)={\rm res}_\lambda\left(\det\left(  i\mathcal{M}\Bigl(\pm i b,N_2-N_1-\frac{3}{2},N_2\Bigr)+\lambda\right), \frac{\partial }{\partial \lambda} \det\left(  i\mathcal{M}\Bigl(\pm i b,N_2-N_1-\frac{3}{2},N_2\Bigr)+\lambda\right)\right),
\end{aligned}
$$
and
$$
\begin{aligned}
&\frac{H_{N_2+1,N_1+1}(a)}{c_{(N_2+2),(N_1+1)}}=
\\
&={\rm res}_\lambda\left(\det\left(\mathcal{M}\Bigl(b,N_1-N_2-\frac{3}{2},N_1\Bigr)+\lambda\right), \det\left(  i\mathcal{M}\Bigl(\pm i b,N_2-N_1-\frac{3}{2},N_2\Bigr)+\lambda\right) \right).
\end{aligned}
$$

     %%%%%%%%%%%%%%%%%%%%%%%%%%%%%%

\section{Appendix 1: factorization of the resultant}

We prove  a  factorization of the resultant, in the following lemma. 

\ble
\label{1febbraio2026-5}

Consider an $(n+m)\times (n+m)$  square matrix with block partition 
$$M=\left(\begin{array}{c|c} M_n & \boldsymbol{0}_{nm}
\\
\hline
N_{mn} & M_m
\end{array}
\right),
$$
where $M_n$ is an $n\times n$ matrix, $M_m$ is an $m\times m$ matrix, and $N_{mn} $ is an $m\times n$ matrix, while $  \boldsymbol{0}_{nm}$ is the $n\times m$ matrix with zero entries. 
Let 
$$
p(\lambda):=\det(M-\lambda),\quad p_n(\lambda):=\det(M_n-\lambda),\quad p_m(\lambda):=\det(M_m-\lambda),
$$
and consider the resultants with respect to $\lambda$:
$$ 
 r_{nm}:=\mathrm{res}_\lambda(p_n(\lambda),p_m(\lambda))
,
\quad\quad
r_j:=\mathrm{res}_\lambda\left(p_j(\lambda),\frac{\partial  p_j(\lambda)}{\partial \lambda}\right),\quad j=n,m.
$$
Then, the following factorization holds
$$ 
\mathrm{res}_\lambda\left(p(\lambda),\frac{\partial p(\lambda)}{\partial \lambda} \right)= (-1)^{nm}\,r_n\,r_m\,r_{nm}^2.
$$

\ele 

\begin{proof}
It is straightforward to see that  the characteristic polynomial  of $M$  factorizes as 
$$ 
p(\lambda)=p_n(\lambda)\,p_m(\lambda).
$$
From this, we receive
$$
\begin{aligned}
{\rm res}_\lambda\Bigl(p(\lambda),~\frac{\partial p(\lambda)}{\partial \lambda}\Bigr) 
&={\rm res}_\lambda\left(p_n(\lambda)p_m(\lambda),\frac{\partial}{\partial \lambda}\Bigl(p_n(\lambda)p_m(\lambda)\Bigr)
\right)
\\
&=    {\rm res}_\lambda \left(p_n(\lambda),\frac{\partial}{\partial \lambda}\Bigl(p_n(\lambda)p_m(\lambda)\Bigr)
\right)
\cdot  {\rm res}_\lambda \left( p_m(\lambda),\frac{\partial}{\partial \lambda}\Bigl(p_n(\lambda)p_m(\lambda)\Bigr)\right).
\end{aligned}
$$
The last line follows from the properties of the resultant \cite{GKZ}. 
Let us consider the first factor in the last line above, the second being treated in an analogous way.
We have 
$$
  {\rm res}_\lambda \left(p_n,\frac{\partial}{\partial \lambda}\Bigl(p_n p_m \Bigr)\right)
  =
   {\rm res}_\lambda \left(p_n,\frac{\partial p_n}{\partial \lambda}p_m+p_n \frac{\partial p_m}{\partial \lambda}\right)
   .
   $$
We claim that 
$$
 {\rm res}_\lambda \left(p_n,\frac{\partial p_n}{\partial \lambda}p_m+p_n \frac{\partial p_m}{\partial \lambda}\right)= {\rm res}_\lambda \left(p_n,\frac{\partial p_n}{\partial \lambda}p_m\right).
 $$
 In order  to prove this, we recall the Bezout formula \cite{GKZ}. Consider two polynomials of degrees $r$ and $s$ respectively:
 $$ 
 f(\lambda)=a_r\lambda^r+a_{r-1}\lambda^{r-1}+\dots+a_0,\quad \hbox {with }a_0=1,
 $$ 
 
 $$ 
 g(\lambda)=b_s\lambda^s+b_{s-1}\lambda^{s-1}+\dots+ b_0,\quad \hbox {with } s\geq r+1.
 $$
Then, Bezout formula says that
$$ 
{\rm res}_\lambda(f,g)=\det 
\begin{pmatrix}
c_s & c_{s+1} & \cdots & c_{s+r-1}
\\
c_{s-1} & c_s & \cdots & c_{s+r-2}
\\
\vdots &\vdots &\ddots & \vdots 
\\
c_{s-r+1} & c_{s-r+2} &  \cdots & c_s
\end{pmatrix},
$$
where 
$$ 
\frac{g(\lambda)}{f(\lambda)}
= c_0+c_1\lambda+c_2\lambda^2+\dots ~=\sum_{\ell=0}^\infty c_\ell \lambda^\ell,
$$ 
is the Taylor expansion at $\lambda=0$. If $a_0\neq 0$ is not equal to 1, then 
$$ 
{\rm res}_\lambda(f,g)= a_0^s\cdot  {\rm res}_\lambda(\widetilde{f},g),\quad \quad \widetilde{f}:= \frac{f}{a_0},
$$
so that we can proceed with the Bezout formula for ${\rm res}_\lambda(\widetilde{f},g)$.

\vskip 0.2 cm 
In our case, first consider the case when $p_n(0)\neq 0$. Let 
$$ 
f(\lambda):= \frac{p_n(\lambda)}{p_n(0)}.
$$
We have 
$$
 {\rm res}_\lambda \left(p_n,\frac{\partial p_n}{\partial \lambda}p_m+p_n \frac{\partial p_m}{\partial \lambda}\right)=
 p_n(0)^{m+n-1}
 \cdot  {\rm res}_\lambda \left(f,\frac{\partial f}{\partial \lambda}p_m+f\frac{\partial p_m}{\partial \lambda}\right).
 $$
 We apply Bezout formula to $f(\lambda)$ and $g(\lambda)$, where 
$$
 g(\lambda):= \frac{\partial f(\lambda)}{\partial \lambda}p_m(\lambda)+f(\lambda)\frac{\partial p_m(\lambda)}{\partial \lambda}.
$$
Let us write 
$$ 
\begin{aligned}
&f(\lambda)=a_n\lambda^n+a_{n-1}\lambda^{n-1}+\dots +1,
\\
&p_m(\lambda)=\beta_m\lambda^m+\beta_{m-1}\lambda^{m-1}+\dots + \beta_0.
\end{aligned}
$$
We need some of the coefficients $c_\ell$ of the Taylor expansion at $\lambda=0$ of the ratio $g/f$, namely of 
$$ 
\underbrace{\frac{1}{f}\frac{\partial f}{\partial \lambda} p_m}_{(1)}+\underbrace{\frac{\partial p_m}{\partial \lambda}}_{(2)}= \underbrace{
\frac{(na_n\lambda^{n-1}+\dots + a_1)(\beta_m\lambda^m+\dots +\beta_0)}{a_n\lambda^n+\dots +1}}_{(1)}+\underbrace{m\beta_m\lambda^{m-1}+\cdots +\beta_1}_{(2)}.
$$
Notice that in this case the degrees in Bezout formula are 
$$ 
r=n, \quad s=m+n-1.
$$ 
We need the coefficients $c_\ell$ for 
$$
s-r+1\leq \ell \leq s+r-1,\quad \hbox{ that is } \quad m\leq \ell \leq m+2n-2.
$$
The term (2) contributes up to  $c_0$, $c_1$, ..., $c_{m-1}$. Only   (1) contributes to $c_m\lambda^m+\dots + c_{m+2n-2}\lambda^{m+2n-2}$. This and Bezout formula prove that 
$$ 
{\rm res}_\lambda \left(f,\frac{\partial f}{\partial \lambda}p_m+f\frac{\partial p_m}{\partial \lambda}\right)= 
{\rm res}_\lambda \left(f,\frac{\partial f}{\partial \lambda}p_m\right)
.
$$
The claim is proved if $p_n(0)\neq 0$. If $p_n(0)=0$,  let 
$$
\widetilde{p}_n(\lambda):=p_n(\lambda)+a_0,\quad a_0\neq 0.
$$
Then, for $\widetilde{p}_n$ the proof above gives the equality 
$$
 {\rm res}_\lambda \left(\widetilde{p}_n,\frac{\partial \widetilde{p}_n}{\partial \lambda}p_m+\widetilde{p}_n \frac{\partial p_m}{\partial \lambda}\right)= {\rm res}_\lambda \left(\widetilde{p}_n,\frac{\partial \widetilde{p}_n}{\partial \lambda}p_m\right).
 $$
Both sides of the formula above are polynomial of degree $s=m+n-1$ in $a_0$, so the equality holds in the limit $a_0\to0$. As a result, the claim is proved also in the case $p_n(0)=0$. 

\vskip 0.2 cm 
Putting all together, we receive
$$ 
\begin{aligned}
{\rm res}_\lambda \left(
p(\lambda),\frac{\partial p(\lambda)}{\partial \lambda}\right)
&=
{\rm res}_\lambda \left( p_n,\frac{\partial p_np_m}{\partial \lambda}\right){\rm res}_\lambda \left( p_m,\frac{\partial p_np_m}{\partial \lambda}\right)
\\
& 
= 
{\rm res}_\lambda \left( p_n,\frac{\partial p_n}{\partial \lambda}p_m\right){\rm res}_\lambda \left( p_m,\frac{\partial p_m}{\partial \lambda}p_n\right)
\\
&
=(-1)^{mn}
{\rm res}_\lambda \left( p_n,\frac{\partial p_n}{\partial \lambda}\right)
\cdot
{\rm res}_\lambda \left( p_m,\frac{\partial p_m}{\partial \lambda}\right)
\cdot 
\left({\rm res}_\lambda \left( p_n,p_m\right)\right)^2.
\end{aligned}
$$

\end{proof}

  %%%%%%%%%%%%%%%%%%%%%%%%%%%%%%

\section{Appendix 2: Non-homogeneous equation in general, another proof of Lemmas \ref{3giugno2026-2} and \ref{3giugno2026-7}}

      \label{7maggio2026-5}
   
  We rewrite \eqref{30aprile2024-2} as    \be
\label{7maggio2026-2}
\frac{d^2y}{dx^2}+(\lambda-V(x))y=0,
\ee
where 
$$
V(x)=V(x,b,M,\gamma):=x^6+2bx^4+(b^2-2M-3)x^2+\frac{\gamma}{x^2}.
$$
Let $y(x,\lambda)$ be a solution holomorphic on $\mathcal{R}\times\mathbb{C}$. 
   Then, consider the non-homogeneous equation
   $$
   \frac{d^2 w}{dx^2}+\left(\lambda-V(x)\right)w= y(x,\lambda).
    $$
 Differentiating \eqref{7maggio2026-2} with respect to $\lambda$ we receive
    $$
 \frac{d^2}{dx^2}   \left(-\frac{\partial}{\partial\lambda}y(x,\lambda) \right)+(\lambda-V(x))\, \left(-\frac{\partial}{\partial\lambda}y(x,\lambda) \right)=y(x,\lambda).
    $$
    Therefore, the general solution of the non-homogeneous equation is 
\be
\label{7maggio2026-8}
    w(x;\lambda)=     -\frac{\partial}{\partial\lambda}y(x,\lambda) +\,\underbrace{k(\lambda)\, y(x,\lambda)+ h(\lambda)\,  y(x,\lambda)\int_{x_o}^x \frac{ds}{ y(s,\lambda)^2}}_{\hbox{general sol. of  \eqref{7maggio2026-2}}},
    \ee
    where $k(\lambda)$ and $h(\lambda)$ are arbitrary holomorphic functions.
    
    \vskip 0.3 cm 
    
    We can rewrite the general solution in two specific cases.     
     \vskip 0.2 cm 
   1)  If  $y(x,\lambda)=y_1^{(\nu)}(x,\lambda)$, then the general solution of the non-homogeneous equation is
     \be
     \label{7maggio2026-8-2}
    w_1^{(\nu)}(x;\lambda)=     -\frac{\partial}{\partial\lambda}y_1^{(\nu)}(x,\lambda) +
    \,\underbrace{\left(k(\lambda)\, y_1^{(\nu)}(x,\lambda)+ h(\lambda)\,  y_2^{(\nu)}(x,\lambda)\right)}_{\hbox{general sol. of  \eqref{7maggio2026-2}}}.
    \ee
  For $x\to\infty$ in $\mathcal{S}_{2\nu-1}\cup \mathcal{S}_{2\nu}$ we have  
     $$ -\frac{\partial}{\partial\lambda}y_1^{(\nu)}(x,\lambda) \sim \sum_{k=0}^\infty w_{2k}(\lambda)
      x^{-2k}\,x^M\,
     e^{-\vartheta(x)}, \quad   w_{2k}(\lambda):=-\frac{\partial }{\partial\lambda} c_{2k}(\lambda)
     $$ 
     where $ c_{2k}(\lambda) $ is \eqref{6maggio2026-1}. 
If follows that  $ w_1^{(\nu)}(x;\lambda)$ behaves at infinity  as a series in $x^{-2k}$ times the factor $x^M \,\exp\{-\vartheta(x)\}$ if and only if $h(\lambda)=0$. 
The factorization \eqref{3giugno2026-5} implies that  
$$
w_{2k}(\lambda)=\frac{c_0(-1)^{k+1}}{k!\,4^k}\,\left(
\frac{\partial \chi_{N+1}(\lambda)}{\partial\lambda}\, \mathfrak{X}_{k-N-1}(\lambda)+\chi_{N+1}(\lambda)\,\frac{\partial\mathfrak{X}_{k-N-1}(\lambda)}{\partial\lambda}
\right),\quad k\geq N+1. 
$$
Hence, if $\Lambda$ is a roof  of $\chi_{N+1}(\lambda)$, we have $y_1^{(\nu)}(x,\Lambda) =y_1(x,\Lambda)$, and if the  algebraic multiplicity is $\geq 2$, we further have 
$$\left.\frac{\partial\chi_{N+1}(\lambda)}{\partial\lambda}\right|_{\lambda=\Lambda}=0,
\quad\Longrightarrow\quad 
w_{2k}(\Lambda)=0\quad \forall\, k\geq N+1.
$$
This is another proof of Lemma \ref{3giugno2026-2}.
    \vskip 0.2 cm 
    2) If  $y(x,\lambda)=y_2^{(\nu)}(x,\lambda)$, then the general solution of the non-homogeneous equation is
 \be
     \label{7maggio2026-8-3}
    w_2^{(\nu)}(x;\lambda)=     -\frac{\partial}{\partial\lambda}y_2^{(\nu)}(x,\lambda) +\,\left(k(\lambda)\, y_1^{(\nu)}(x,\lambda)+ h(\lambda)\,  y_2^{(\nu)}(x,\lambda)\right),
\ee
    For $x\to\infty$ in $\mathcal{S}_{2\nu}\cup \mathcal{S}_{2\nu+1}$, we have 
     $$ -\frac{\partial}{\partial\lambda}y_2^{(\nu)}(x,\lambda) \sim \sum_{k=0}^\infty f_{2k}(\lambda)
      x^{-2k}\,x^{-M-3}\,
     e^{\vartheta(x)}, \quad   f_{2k}(\lambda)=-\frac{\partial}{\partial\lambda} \,d_{2k}(\lambda)
     $$ 
     where $ d_{2k}(\lambda) $ is \eqref{6maggio2026-3-bis}. 
     In this case,  $ w_2^{(\nu)}(x;\lambda)$ behaves at infinity  as a series in $x^{-2k}$ times the factor $x^{-M-3} \,\exp\{\vartheta(x)\}$ if and only if $k(\lambda)=0$. 
  The factorization    \eqref{3giugno2026-6} implies that 
  $$
  f_{2k}(\lambda)= -\frac{d_0}{k!\,4^k}\, \left( \frac{\partial\widetilde{\chi}_{N+1}(\lambda)}{\partial\lambda} \, \widetilde{\mathfrak{X}}_{k-N-1}(\lambda)
  +\widetilde{\chi}_{N+1}(\lambda) \,\frac{ \partial \widetilde{\mathfrak{X}}_{k-N-1}(\lambda)}{\partial\lambda} \right)
     $$
     Hence, if $\Lambda$ is a roof  of $\widetilde{\chi}_{N+1}(\lambda)$, we have $y_2^{(\nu)}(x,\Lambda) =y_2(x,\Lambda)$, and if the  algebraic multiplicity is $\geq 2$, we further have 
$$\left.\ \frac{\partial\widetilde{\chi}_{N+1}(\lambda)}{\partial\lambda}\right|_{\lambda=\Lambda}=0,
\quad\Longrightarrow\quad 
f_{2k}(\Lambda)=0\quad \forall\, k\geq N+1.
$$
This is another proof of Lemma \ref{3giugno2026-7}.


\begin{thebibliography}{99} 

\bibitem{BenD-1} C. Bender and G. Dunne. {\it Quasi- Exactly Solvable Systems and Orthogonal Polynomials.} J. Math. Phys. 37 (1996), 6-11.

\bibitem{BenD-2} C. Bender, G. Dunne, and M. Moshe. {\it Semiclassical Analysis of Quasi-exact Solvability.}  Phys. Rev. A 55:2 (1997), 2625-2629.

\bibitem{BCG-1}  M. Bertola, E.E. Chavez Heredia, T. Grava,: {\it  Exactly Solvable Anharmonic Oscillator, Degenerate Orthogonal Polynomials and Painlev\'e II}. Communications in Mathematical Physics, 405(2), 52 (2024).

\bibitem{BazZ} V. Bazhanov, S. Lukyanov, A. Zamolodchikov: {\it 
Spectral determinants for Schr\"odinger equation and Q-operators of conformal field theory},
J. Stat. Phys. 102 (2001) 567-576.


\bibitem{Maso-Bridgl-1}
T. Bridgeland, D. Masoero: {\it On the monodromy of the deformed cubic oscillator}. Math. Ann. 385, 193-258 (2023). https://doi.org/10.1007/s00208-021-02337-w

\bibitem{BM} R.J. Buckingham, P.D. Miller: {\it Large-Degree Asymptotics of Rational Painlev\'e-IV Solutions by the Isomonodromy Method}, Constructive Approximation (2022) 56:233-443.

\bibitem{CGM2023} G.\,Cotti, D.\,Guzzetti, D.\,Masoero: {\it Asymptotic solutions for linear ODEs with not-necessarily meromorphic coefficients: a Levinson type theorem on complex domains, and applications}, Journal of Differential Equations {\bf 428}, (2025), 1-58. 


\bibitem{Deg1} G. Degano: {\it ODE/IM Correspondence in the Semiclassical Limit: Large Degree Asymptotics of the Spectral Determinants for the Ground State Potential}, Constr Approx (2026). https://doi.org/10.1007/s00365-026-09750-x

\bibitem{Deg2} G. Degano, D. Masoero :{\it A primer of the complex WKB method, with application to the ODE/IM correspondence }, (2025), arXiv:2501.05957 

\bibitem{D1..?} P. \,Dorey, R. \,Tateo:
{\it Anharmonic oscillators, the thermodynamic Bethe ansatz, and nonlinear integral equations},
J. Phys. A 32 (1999) L419-L425.

\bibitem{DDT} P.\,Dorey, C.\,Dunning, and R.\,Tateo, {\it The ODE/IM Correspondence}, J.\,Phys.\,A Math.\,Theor. 40(32), 1 (2007).

\bibitem{Erem-Shap} A. Eremenko, A. Gabrielov, B. Shapiro: {\it Zeros of eigenfunctions of some anharmonic oscillators}. Annales de l'Institut Fourier,  58 (2008), 603-624. 

\bibitem{Gant} F.R. Gantmacher: {\it The Theory of Matrices}, AMS Chelsea Publishing (1958-2000). 

\bibitem{GKZ}  I.M. Gelfand, M.M. Kapranov, A.V. Zelevinsky: {\it Discriminants, Resultants and Multidimensional Determinants}, Birkh\"auser 1994.

\bibitem{JM-2} M. Jimbo, T. Miwa: {\it  Monodromy Preserving Deformations of Linear Ordinary Differential Equations with Rational Coefficients (II)}. Physica , D 2 , (1981), 407-448

\bibitem {DMTritr-1} D. Masoero: {\it Poles of int\'egrale tritronqu\'ee and anharmonic oscillators. A WKB approach}, 
J. Phys. A 43 (2010), no. 9, 095201, 28 pp.


 \bibitem {DMTritr-2} D. Masoero:   {\it Poles of int\'egrale tritronqu\'ee and anharmonic oscillators. Asymptotic localization from WKB analysis}, 
Masoero, Davide 
Nonlinearity 23 (2010), no. 10, 2501-2507.

\bibitem{MRV} D.\,Masoero, A.\,Raimondo, and D.\,Valeri, {\it Bethe Ansatz and the Spectral Theory of affine Lie algebra-valued connections I. The simply-laced case}, Commun.\,Math.\,Phys. 344(3), 719--750 (2016).

\bibitem{MRV2} D.\,Masoero, A.\,Raimondo, and D.\,Valeri,
{\it Bethe Ansatz and the Spectral Theory of affine Lie algebra--valued connections II. The non simply--laced case}, Commun.\,Math.\,Phys. 349(3), 1063--1105 (2017).


\bibitem{MR} D. Masoero, P. Roffelsen: {\it Poles of Painlev\'e IV Rationals and their Distribution}, SIGMA 14 (2018), 002, 49 pages.

\bibitem{MR1} D. Masoero, P. Roffelsen: {\it Roots of generalised Hermite polynomials when both parameters are large}, Nonlinearity 34 (2021), 1663-1732.


\bibitem{NY} M. Noumi, Y. Yamada: {\it Symmetries in the fourth Painlev\' equation and Okamoto polynomials} .Nagoya
Math. J. 153, 53-86 (1999).

\bibitem{Singh} V. Singh, S. N. Biswas, and K. Datta. {\it Anharmonic Oscillator and the Analytic Theory of Continued Fractions.} Phys. Rev. D18 (1978), 1901-1908. 

\bibitem{ST-1} B. Shapiro, M. Tater: {\it On spectral asymptotic of quasi-exactly solvable quartic potential}, Anal.
Math. Phys. 12 (2022), no. 1, Paper no. 2, 35 pp.


\bibitem{ST-2} B. Shapiro, M. Tater: {\it Asymptotics and Monodromy of the Algebraic Spectrum of Quasi-Exactly Solvable Sextic Oscillator},  Experimental Mathematics,  (2017), 16-23. DOI: 10.1080/10586458.2017.1325792


\bibitem{ST} M.A.  Shifman, A.V. Turbiner {\it  Quantal problems with partial algebraization of the spectrum}. Commun.Math. Phys. 126, 347-365 (1989). https://doi.org/10.1007/BF02125129


\bibitem{Suzz} J. Suzuki: {\it 
Anharmonic oscillators, spectral determinant and short exact sequence of $U_q(\widehat{sl}_2)$},
J. Phys. A 32 (1999) 183-188
\bibitem{Sh4}  Y. Sibuya: {\it Simplification of a System of Linear Ordinary Differential Equations about a Singular Point}, Funkcial. Ekvac, {\bf 4} (1962), 29-56.



\bibitem{Sh2}  Y. Sibuya: {\it Perturbation of Linear Ordinary Differential Equations  at Irregular Singular Points}, Funkcial. Ekvac, {\bf 11} (1968), 235-146.


\bibitem{Turb-1} A. Turbiner: {\it Quasi-exactly Solvable Problems and  sl(2) Algebra.} Comm. Math. Phys. 118 (1988), 467- 474.

\bibitem{Turb-2} A. Turbiner and A. Ushveridze. {\it Spectral Singularities and the Quasi Exactly Solvable Problem}.  Phys. Lett. 126A (1987), 181-183. 




\bibitem{Turb-1} A. Turbiner: {\it One-dimensional quasi-exactly solvable Schr\"odinger equations}, Physics Reports 642 (2026), 1-71. 

\bibitem{Ush} A. Ushveridze. {\it Quasi-exactly Solvable Models in Quantum Mechanics}. Bristol: Institute of Physics Publish- ing, 1994. xiv+465 pp.


\bibitem{Wasow} W Wasow: {\it Asymptotic Expansions for Ordinary Differential Equations}. Dover (1965)

\end{thebibliography}
\end{document}